%% file: ARXIV-verification-paper.tex
\documentclass[sigconf]{acmart}

\acmConference[]{Conference 2025}{}{}
\usepackage[T1]{fontenc}
\usepackage{graphicx}
\usepackage{multirow}
\usepackage{color}
\usepackage{listings}
\usepackage{wrapfig}
\usepackage{soul}
\usepackage{amsmath}
\usepackage{tikz}
\usepackage{array}
\usepackage{circuitikz}
\usepackage{fancyvrb}
\usepackage{subcaption}
\usepackage{algorithm}
\usepackage{algpseudocode}
\usepackage{algorithmicx}

\newtheoremstyle{acmexample} % Define a new style
  {2mm}   % Space above
  {\topsep}   % Space below
  {\normalfont} % Body font (non-italic)
  {}          % Indent amount
  {\bfseries} % Theorem head font
  {.}         % Punctuation after theorem head
  { }         % Space after theorem head
  {}          % Theorem head spec (can be left empty, as it is in acmart)

\theoremstyle{acmexample} % Use the new style

\newtheorem{example}{Example}
\newtheorem{definition}{Definition}

\newtheoremstyle{acmtheorem} % Define a new style
  {2mm}   % Space above
  {1mm}   % Space below
  {\itshape} % Body font (non-italic)
  {}          % Indent amount
  {\bfseries\scshape} % Theorem head font
  {.}         % Punctuation after theorem head
  { }         % Space after theorem head
  {}          % Theorem head spec (can be left empty, as it is in acmart)

\theoremstyle{acmtheorem} % Use the new style
\newtheorem{theorem}{Theorem}

\newcolumntype{P}[1]{>{\raggedright\arraybackslash}p{#1}}
\usepackage{booktabs}

\definecolor{prismgreen}{rgb}{0, 0.6, 0}
\newcommand{\prismfont}{\scriptsize\selectfont}

\lstdefinelanguage{Prism}{ % syntax highlight via font
basicstyle=\color{red}\prismfont\sffamily, % small true type font (like courier)
keywords={bool,C,ceil,const,ctmc,double,dtmc,endinit,endmodule,endrewards,endsystem,F,false,floor,formula,G,global,I,init,int,label,max,mdp,min,module,nondeterministic,P,Pmin,Pmax,prob,probabilistic,R,rate,rewards,Rmin,Rmax,S,stochastic,system,true,U,X,observables,endobservables},
keywordstyle={\bfseries\color{black}},
numberstyle=\tiny\color{black},
comment=[l] {//}, morecomment=[s]{/*}{*/}, % single and multi-line
commentstyle= \color{prismgreen}, % dark green
tabsize=4, % tab treatment (going to be fixed in Prism)
captionpos=b, % put captions at the bottom
escapechar=@, % write LaTeX comments escaped by @ symbol
literate=%
        {-}{{\textcolor{black}{$-$}}}{1}%
        {->}{{\textcolor{black}{$\rightarrow{}$}}}{2}%
        {0}{{\textcolor{blue}{0}}}{1}%
             {1}{{\textcolor{blue}{1}}}{1}%
             {p1}{{\textcolor{red}{p1}}}{1}%
             {2}{{\textcolor{blue}{2}}}{1}%
             {p2}{{\textcolor{red}{p2}}}{1}%
             {3}{{\textcolor{blue}{3}}}{1}%
             {4}{{\textcolor{blue}{4}}}{1}%
             {5}{{\textcolor{blue}{5}}}{1}%
             {6}{{\textcolor{blue}{6}}}{1}%
             {7}{{\textcolor{blue}{7}}}{1}%
             {8}{{\textcolor{blue}{8}}}{1}%
             {9}{{\textcolor{blue}{9}}}{1}%
             {.}{{\textcolor{blue}{.}}}{1}%
             {=}{{\textcolor{black}{=}}}{1}%
             {[}{{\textcolor{black}{[}}}{1}%
             {]}{{\textcolor{black}{]}}}{1}%
             {;}{{\textcolor{black}{;}}}{1}%
             {+}{{\textcolor{black}{+}}}{1}%
             {*}{{\textcolor{black}{*}}}{1}%
             {:}{{\textcolor{black}{:}}}{1}%
             {\&}{{\textcolor{black}{\&}}}{1}%
             {|}{{\textcolor{black}{|}}}{1}%
             {?}{{\textcolor{black}{?}}}{1}%            
             {"}{{\textcolor{black}{"}}}{1}%            
             {(}{{\textcolor{black}{(}}}{1}%
             {)}{{\textcolor{black}{)}}}{1}%
             {'}{{\textcolor{black}{'}}}{1}%
}

\newcommand{\squishlist}{
   \begin{list}{$\bullet$}
    { \setlength{\itemsep}{2pt}    \setlength{\parsep}{0pt}
      \setlength{\topsep}{5pt}     \setlength{\partopsep}{0pt}
      \setlength{\leftmargin}{2.5em} \setlength{\labelwidth}{1em}
      \setlength{\labelsep}{0.5em} } }

\newcommand{\squishlisttwo}{
   \begin{list}{$\bullet$}
    { \setlength{\itemsep}{2pt}    \setlength{\parsep}{0pt}
      \setlength{\topsep}{5pt}     \setlength{\partopsep}{0pt}
      \setlength{\leftmargin}{1.35em} \setlength{\labelwidth}{1em}
      \setlength{\labelsep}{0.5em} } }

\newcommand{\squishend}{
    \end{list}  }

\setcopyright{none} % to remove the copyright notice
\renewcommand\footnotetextcopyrightpermission[1]{}

\begin{document}

\title{Verification and External Parameter Inference\\ for Stochastic World Models}

\author{Radu Calinescu, Sinem Getir Yaman, Simos Gerasimou, Gricel V\'{a}zquez and Micah Bassett}
 \affiliation{%
   \institution{Department of Computer Science, University of York}
   \city{York}
   \country{United Kingdom}}
\email{{radu.calinescu,sinem.getir.yaman,simos.gerasimou,gricel.vazquez,micah.bassett}@york.ac.uk}

\renewcommand{\shortauthors}{Calinescu et al.}

\begin{abstract}
  Given its ability to analyse stochastic models ranging from discrete and continuous-time Markov chains to Markov decision processes and stochastic games, probabilistic model checking (PMC) is widely used to verify system dependability and performance properties. However, modelling the behaviour of, and verifying these properties for many software-intensive systems requires the joint analysis of multiple interdependent stochastic models of different types, which existing PMC techniques and tools cannot handle. To address this limitation, we introduce a tool-supported \emph{UniversaL stochasTIc Modelling, verificAtion and synThEsis} (ULTIMATE) framework that supports the representation, verification and synthesis of heterogeneous multi-model stochastic systems with complex model interdependencies. Through its unique integration of multiple PMC paradigms, and underpinned by a novel verification method for handling model interdependencies, ULTIMATE unifies---for the first time---the modelling of probabilistic and nondeterministic uncertainty, discrete and continuous time, partial observability, and the use of both Bayesian and frequentist inference to exploit domain knowledge and data about the modelled system and its context. A comprehensive suite of case studies and experiments confirm the generality and effectiveness of our novel verification framework. 
\end{abstract}

\maketitle
\pagestyle{plain} % add this after maketitle to remove running header

\section{Introduction}

Software-intensive systems of all types, from simple computer applications and complex cyber-physical systems to sophisticated AI agents, operate under uncertainty. This uncertainty stems from factors including the \emph{nondeterminism} inherent in their user inputs and the availability of multiple system actions to select from, the \emph{stochasticity} of the execution times and effects of the selected actions, and the \emph{partial observability} resulting from their use of never-perfect machine learning components to perceive the environment. To consider these factors when verifying the dependability, performance and other quality properties of such systems, software engineers are often resorting to \emph{probabilistic model checking} (PMC)~\cite{katoen2016probabilistic,KNP17}.

There are many reasons for this frequent use of PMC for the formal modelling and verification of software-intensive systems, e.g.~\cite{epifani2009model,DBLP:journals/tse/CalinescuGKMT11,DBLP:journals/tse/CalinescuWGIHK18,DBLP:journals/tse/PatersonC20,DBLP:conf/icse/CalinescuK09,ghezzi2013model}. The models supported by PMC (discrete- and continuous-time Markov chains, Markov decision processes, partially observable Markov decision processes, etc.) capture key aspects of the uncertainty affecting such systems. The use of expressive probabilistic temporal logics~\cite{hansson-jonsson1994,Andova-etal2004,aziz1996,de1998formal,chen2013automatic} allows the specification of a wide range of properties over these models. The development of efficient algorithms and probabilistic model checkers such as PRISM~\cite{KNP11} and Storm~\cite{hensel2022probabilistic} for the verification of these properties have greatly eased the adoption of PMC across application domains~\cite{10.1145/2933575.2934574,KNP17,KNP22}. The emergence of parametric model checking for Markov chains with transition probabilities and rewards specified as parameters~\cite{Daws:2004:SPM:2102873.2102899,param,DBLP:journals/tse/FangCGA23,gainer2018accelerated} supports the synthesis of probabilistic models~\cite{DBLP:journals/jss/CalinescuCGKP18,vcevska2019shepherding,DBLP:journals/ase/GerasimouCT18,junges2024parameter} corresponding to software system designs~\cite{DBLP:journals/jss/CalinescuCGKP18,DBLP:conf/cav/PasareanuMGYICY23} and discrete-event software controllers~\cite{10496502} guaranteed to meet complex sets of requirements. Last but not least, the integration of PMC with frequentist~\cite{DBLP:journals/jss/AlasmariCPM22,DBLP:journals/tr/CalinescuGJPRT16} and Bayesian~\cite{epifani2009model,DBLP:conf/kbse/ZhaoCGRF20,zhao2024bayesian} inference enables the exploitation of expert knowledge and of data from logs and runtime monitoring to improve the accuracy of probabilistic models, and thus the validity of their verification.

This richness of the PMC landscape~\cite{katoen2016probabilistic,Hensel2025} enables the verification of key quality properties for numerous software-intensive systems. Nevertheless, analysing all relevant properties of many complex systems requires the \emph{joint use} of several of these PMC modelling, verification and synthesis methods. These systems comprise interacting components that cannot be verified entirely independently, and that exhibit a combination of discrete and continuous stochastic behaviour, nondeterminism, partial observability, etc. Despite notable research on the assume-guarantee verification of interdependent models of the same type with simple model interdependencies~\cite{calinescu2012compositional,feng2011learning,kwiatkowska2010assume}, the PMC of more general types of multi-model stochastic systems is currently underexplored. 

\begin{figure}[b]
    \centering
    \includegraphics[width=\linewidth]{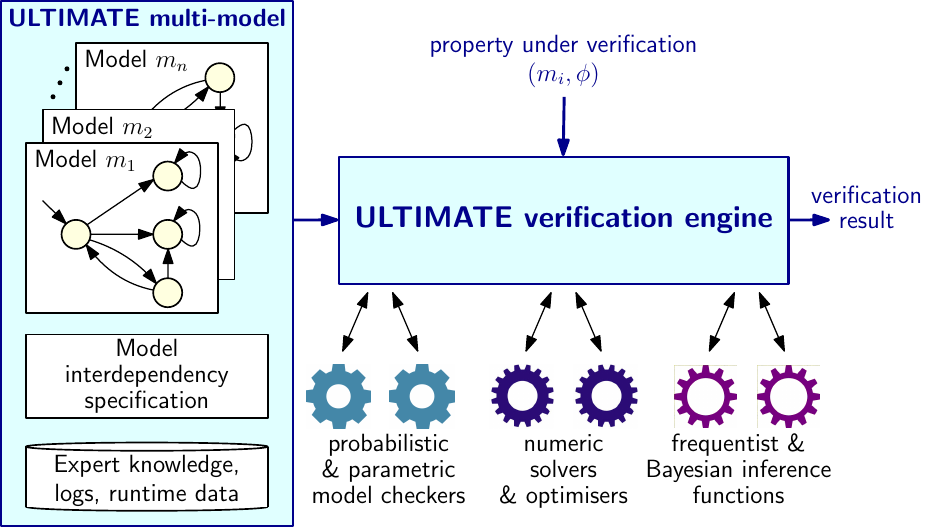}
    \caption{ULTIMATE multi-model verification}
    \label{fig:approach}
\end{figure}

To address this gap, we introduce a \emph{UniversaL stochasTIc Modelling, verificAtion and synThEsis} (ULTIMATE) framework that supports the representation, verification and (when the selection of system actions is required) synthesis of heterogeneous multi-model stochastic systems with complex model interdependencies. As shown in Figure~\ref{fig:approach}, the ULTIMATE verification engine at the core of our framework takes two inputs. The first input, called an \emph{ULTIMATE multi-model}, is a set of $n>1$ stochastic models of the types encountered in PMC, together with (i)~a formal specification of their interdependencies, and (ii)~expert knowledge, logs and runtime data to be used for the estimation of their external parameters. The second input is a formally specified property $\phi$ of one of these models, $m_i$, that needs to be verified by appropriately resolving all relevant model interdependencies, estimating the required parameters, etc. Given these inputs, the ULTIMATE verification engine produces the required verification result by (i)~performing a dependency analysis of the multi-model under verification, (ii)~synthesising the sequence of model analysis and parameter computation tasks required to verify the property, and (iii)~invoking the combination of probabilistic and parametric model checkers, numeric solvers and optimisers, and frequentist and Bayesian inference functions needed to execute these tasks.
Through its unique integration of multiple PMC paradigms, our ULTIMATE framework unifies---for the first time---the modelling of probabilistic and nondeterministic uncertainty, discrete and continuous time, partial observability, and the use of both Bayesian and frequentist inference to exploit domain knowledge and data about the modelled context. 

The main contributions of our paper are:\\[-4.5mm]
\squishlisttwo
\item[1)] A theoretical foundation comprising a formal definition of multi-model stochastic systems, and a novel algorithm for verifying such systems through analysing their constituent models and subsets of models in an order and through methods that consider their interdependencies and co-dependencies;
\item[2)] Tool support that implements our theoretical foundation, automating the verification of multi-model stochastic systems;
\item[3)] A suite of case studies that spans multiple application domains and types of software-intensive systems, and demonstrates the applicability and versatility of our framework.
\squishend

We organised the paper as follows. Section~\ref{sec:background} summarises the types of stochastic models that can be included into an ULTIMATE multi-model, the probabilistic temporal logics used to specify their interdependencies and properties, and the high-level modelling language adopted by our framework. Section~\ref{sec:motivating} presents a motivating example, Section~\ref{sec:foundation} covers our theoretical foundation, and Section~\ref{sec:implementation} introduces the verification tool we implemented to automate the use of the framework. We then describe our case studies and experiments in Section~\ref{sec:evaluation}, compare ULTIMATE to related research in Section~\ref{sec:related}, and conclude the paper with a brief summary in Section~\ref{sec:conclusion}.

\section{Background} \label{sec:background}

\textbf{Stochastic models.} Probabilistic model checking supports the analysis of \emph{stochastic models} comprising \emph{states} that abstract key aspects of the modelled system at different points in time, and \emph{state transitions} that capture its evolution between successive states.\footnote{Ensuring that the model states only capture system aspects relevant to the properties of interest---with other system aspects \emph{abstracted out}---is essential to bound the model size so that its PMC is feasible. For example, the request queue length needs to be captured in the model of a client-server system when analysing the server's response time, while aspects like the server storage space should be abstracted out.} Depending on the model type, the state transitions are taken with probabilities (for \emph{discrete-time models}) or rates (for \emph{continuous-time models}) that reflect the stochastic nature of this evolution. Model states are labelled with \emph{atomic propositions} representing basic properties that hold in those states; and non-negative values termed \emph{rewards} can be assigned to states and transitions. 

When nondeterminism is present, multiple \emph{actions} are possible for at least some states, and each transition is associated with the \emph{action} that enables it. Under \emph{partial observability}, subsets of states are indistinguishable to agents when they select their actions. When modelling multi-agent systems, each action corresponds to a specific agent, or is multi-dimensional and includes a separate action for each agent. Finally, for any state of a discrete-time model, the probabilities of the outgoing transitions associated with the same action (or of all outgoing transitions in the absence of nondeterminism) must add up to 1. 

\begin{table}[t]
\caption{Characteristics of main PMC stochastic models}
\label{table:model-types}
\def\tabcolsep{2pt}
\centering
\sffamily
{\fontsize{8}{9.5}\selectfont 

\vspace*{-2.5mm}
\begin{tabular}{P{2.6cm} P{1.5cm} P{1.3cm} P{1.3cm} P{1cm}}
\toprule
\textbf{Model type} & \textbf{Transitions} & \textbf{Nondeter\-minism} & \textbf{Observa\-bility} & \textbf{\#Agents}  \\ \midrule
discrete-time Markov chain (DTMC) & probabilistic & no & full & 1 \\ \midrule
Markov decision process (MDP) & probabilistic & yes & full & 1  \\  \midrule
probabilistic automaton (PA) & probabilistic & yes & full & 1 \\  \midrule
partially observable MDP (POMDP) & probabilistic & yes & partial & 1\\  \midrule
stochastic game (SG) & probabilistic & yes & full & $2+$ \\ \midrule
continuous-time Markov chain (CTMC) & rate-based & no & full & 1 \\  \bottomrule
\end{tabular}
}

\vspace*{-1mm}
\end{table}

\begin{figure*}
    \centering
\includegraphics[width=\linewidth]{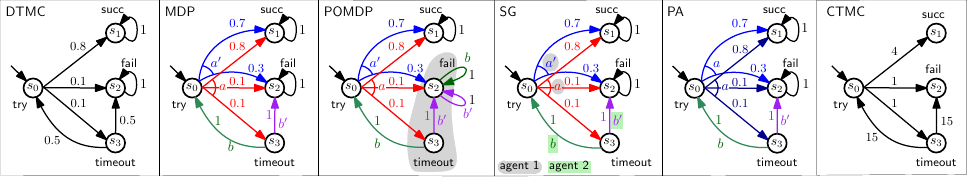}

    \vspace*{-1mm}
    \caption{Examples of stochastic models from Table~\ref{table:model-types}: DTMC modelling an agent's execution of a task which, being tried in state $s_0$, succeeds with probability~0.8 (leading to a DTMC transition to state $s_1$), fails with probability~0.1 (leading to a transition to state $s_2$), or times out with probability 0.1 (yielding a transition to state $s_3$, where the task is re-tried with probability 0.5, or the agent gives up and the task fails with probability 0.5); MDP modelling a variant of the agent in which two actions are available in both state $s_0$ ($a$ and $a'$) and $s_3$ ($b$ and $b'$); POMDP modelling the scenario in which states $s_2$ and $s_3$ are indistinguishable to the agent from the MDP; SG modelling the scenario in which two different agents decide the action selected in state $s_0$ ($a$ or $a'$) and $s_3$ ($b$ or $b'$); PA modelling the presence of two transition probability distributions for action $a$ from state $s_0$; CTMC modelling the rates of transition between the states of the same agent.
    }
    \label{fig:model-type-examples}

    \vspace*{-1mm}
\end{figure*}

Table~\ref{table:model-types} summarises the main types of stochastic models supported by PMC, and a simple example of each model type is shown in Figure~\ref{fig:model-type-examples}. Providing formal definitions of these types of stochastic models is beyond the scope of this paper; such definitions are available, for instance, in~\cite{katoen2016probabilistic,Hensel2025,KNP17}. 

\smallskip\noindent 
\textbf{Property specification.} PMC supports the analysis of stochastic-model properties specified formally in \emph{probabilistic temporal logics}. These expressive logics can be used to encode software-intensive system properties as diverse as `What is the probability that a web server will successfully handle a user request within 2s?', `Will a software controller ensure that a mobile robot will complete its mission without crashing into obstacles with at least $0.995$ probability?' and `What software product line variant can minimise the expected execution time of a workflow (and what is this time)?'

Different logics are suited for each type of stochastic model handled by PMC. Probabilistic computation tree logic (PCTL)~\cite{hansson-jonsson1994} augmented with rewards~\cite{Andova-etal2004} is used to express the properties of discrete-time models such as DTMCs, MDPs, PAs and POMDPs (see Table~\ref{table:model-types} and Figure~\ref{fig:model-type-examples}), while continuous stochastic logic (CSL)~\cite{aziz1996} extends PCTL for continuous-time models such as CTMCs. A range of useful properties of discrete-time models can also be defined in linear temporal logic (LTL)~\cite{pnueli1977temporal}, and PCTL*~\cite{de1998formal}, a temporal logic that combines PCTL and LTL. Finally, the properties of stochastic games are expressed in probabilistic alternating-time temporal logic with rewards (rPATL)~\cite{chen2013automatic}. Again, we are not providing formal definitions of these logics due to space constraints; these definitions are available from the sources cited in this paragraph.

\smallskip\noindent 
\textbf{Model representation.}
Stochastic models are often represented in the high-level modelling language provided by the probabilistic model checker PRISM~\cite{KNP11}---a language that we also adopt in our ULTIMATE framework. 
This language, based on reactive systems~\cite{reactive}, models a system as a parallel composition of multiple \texttt{modules}. Each such module comprises a state space defined by a set of finite-range local variables, and state transitions specified by commands with the generic form:\\[-8.5mm] 

\[ 
[\mathit{action}] \,\,\, \mathit{guard} \to e_1 : \mathit{update}_1 + e_2 : \mathit{update}_2 + \dots + e_m : \mathit{update}_m; \] 

\vspace*{-1mm}\noindent
where:

\vspace*{-1mm}
\squishlisttwo
\item $\mathit{guard}$ is a boolean expression over the variables of all modules; 
\item $e_1$, $e_2$, \ldots, $e_m$ are arithmetic expressions defined over the same variables, and specifying probabilities, $\sum_{i=1}^m e_i = 1$, for discrete-time models, and transition rates for continuous-time models;
\item $\mathit{update}_1$, $\mathit{update}_2$, \ldots, $\mathit{update}_m$ specify changes to the local variables of the module.
\squishend 
If the $\mathit{guard}$ evaluates to true, then $\mathit{update}_i$, $i\in [m]$, is applied with probability $e_i$ for discrete-time models, and with probability $e_i/E$, where $E$ is the sum of all rates associated with true guards within the model, for continuous-time models.

When an $\mathit{action}$ is specified, all modules containing commands with this action must synchronize and execute one of these rules concurrently. Rewards for states and/or transitions can be defined using the \texttt{rewards} \ldots \texttt{endrewards} construct.   

For partially observable models, the observable subsets of states are explicitly defined using the \texttt{observable} construct
\[
  \texttt{observables} \,\,\, v_i,v_j,\ldots \,\,\,\texttt{endobservables}
\]
where $v_i,v_j,\ldots$ are model variables whose values (and associated model states) are observable, with the values of all other model variables being unobservable. State observability affects how the system is verified, particularly when reasoning about the probability of certain outcomes, or synthesising policies of a modelled agent.

Finally, a stochastic game requires the specification of the \emph{players} (i.e., agents) and their control variables using the construct 
\[
\texttt{player}\,\,\, \mathit{pname}\,\,\, \mathit{mname} \,\,\,[a_1,a_2,\ldots,a_n] \,\,\,\texttt{endplayer}.
\]
where $\mathit{pname}$ is the name of a player that executes the module $\mathit{mname}$ and controls the actions $a_1,a_2,\ldots,a_n$.

%\begin{example}
%The DTMC model of a task execution from Figure~\ref{fig:model-type-examples} is encoded in the PRISM language as follows.
%
%\begin{lstlisting}[language={Prism}]
%  dtmc
%
%  module TaskExecution
%    s : [0..3];
%    
%    [] s=0 -> 0.8:(s'=1) + 0.1:(s'=0)+ 0.8:(s'=3);
%    [] s=1 -> 1:(s'=1);
%    [] s=2 -> 1:(s'=2);
%    [] s=3 -> 0.5:(s'=0) + 0.5:(s'=2);
%  endmodule
%\end{lstlisting}
%\end{example}

\section{Motivating example} \label{sec:motivating}

\begin{figure}[b]

\vspace*{-3mm}
    \centering
\includegraphics[width=0.75\linewidth]{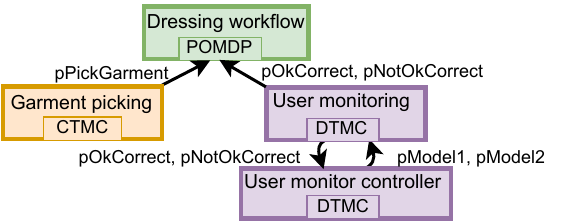}%figures/RAD-motivating-example.pdf}

\vspace*{-2mm}
    \caption{RAD stochastic models and dependencies}\label{fig:RAD-models}
\end{figure}

To motivate the need for the probabilistic model checking of multi-model stochastic systems, we consider the PMC of a robot assistive dressing (RAD) system. This cyber-physical system belongs a domain of growing societal importance due to the significant increase in demand for assistive care driven by an ageing population worldwide~\cite{unfpa2022,who2021}. The RAD system uses a robotic arm to help a person with restricted mobility to dress with a garment such as a coat, enabling them to live independently at home instead of moving into a care home. 
We want to use PMC to synthesise a policy that minimises the system's probability of failing to accomplish the dressing procedure. This requires the joint modelling and analysis of several RAD components and processes:
\squishlist
\item[1.] the picking of the garment by the robot (from a nearby peg);
\item[2.] the monitoring of the user by a deep-learning perception component, to determine whether the user is ok or not;
\item[3.] the control component that configures the user monitor;
\item[4.] the dressing process, which depends on the successful picking of the garment, and must adapt to the (perceived) user state.
\squishend
As shown in Figure~\ref{fig:RAD-models}, we need four interdependent stochastic models of several different types to capture these RAD aspects.

\smallskip
\noindent
\textbf{Garment picking.} One of the essential tasks of the RAD dressing  is the picking of the garment by the robotic arm. To avoid lengthy dressing sessions (which may cause distress to the user), we are interested in the (probability of) successful completion of this task within a 90s time period. A CTMC (which supports the verification of such time-based properties) is therefore used to model the execution of this task. This CTMC (presented in Listing~\ref{lst:RAD-garment-picking-model}) models a robotic arm that tries to pick the garment with a rate \textsf{rPick}, a probability \textsf{psucc} of succeeding in each attempt, and a probability \textsf{pRetry} of retrying the garment picking after an unsuccessful attempt (lines~10--12). If the try is successful, the robotic arm transitions from the initial state \textsf{s=0} to state \textsf{s=1} (labelled with the atomic proposition \textsf{``success''} in line~17). Otherwise, it stays in state \textsf{s=0} for a retry with probability \textsf{pRetry}, or gives up and transition to the fail state \textsf{s=2} with probability $1-\textsf{pRetry}$. 

\begin{lstlisting}[float, belowskip=-1.8\baselineskip, language={Prism}, numbers=left, rulesepcolor=\color{black}, rulecolor=\color{black}, breaklines=true, breakatwhitespace=true, firstnumber=1, firstline=1, 
caption={CTMC modelling the garment picking task},
label={lst:RAD-garment-picking-model}]
ctmc

const double rPick;  // garnment picking rate      
const double psucc;  // successful picking probability
const double pRetry; // retry probability

module pickGarment
  s : [0..2] init 0;              

  [try]  s=0 -> psucc*rPick:(s'=1) + 
                (1-psucc)*rPick*pRetry:(s'=0) +
                (1-psucc)*rPick*(1-pRetry):(s'=2);
  [succ] s=1 -> 1:(s'=1);  
  [fail] s=2 -> 1:(s'=2);                                                         
endmodule

label "success" = s=1;
\end{lstlisting}

The probabilistic model checking of this CTMC can be used to establish the probability \textsf{pPickGarment} of successful garment picking within 90s by verifying the CSL property
\begin{equation}
    \label{eq:RAD-pPickGarment}
     \textsf{pPickGarment} = P_{=?}[F^{[0,90]}\; \textsf{"success"}] 
\end{equation}
We note that the (external) CTMC parameters \textsf{rPick} and \textsf{psucc} (lines 3 and 4) depend on the environment in which the robotic arm is deployed (e.g., on the position of the peg where the garment is located initially, and on the level of lighting from the area). As such, their values need to be derived from experimental data obtained during the preliminary testing of the robotic arm. %, and may be updated as further such data is obtained during its operation.

\smallskip
\noindent
\textbf{User monitoring.} This RAD component uses machine learning (ML) perception to classify the user as content with the ongoing dressing (i.e., \textsf{ok}) or not (\textsf{notok}). The component comprises: 
\squishlisttwo
\item a medium accuracy but fast user-state classifier (ML model~1);
\item a high accuracy, but slower and computationally expensive user-state classifier (ML model~2);
\item a \emph{verifier} that, given an (input, output) pair from model~1, returns \textsf{true} if the output is very likely correct, and \textsf{false} otherwise (see~\cite{10496502} for examples of such verifiers for DNN classifiers).
\squishend
We use the DTMC in Listing~\ref{lst:RAD-user-perception-model} to model the operation of this user monitoring component, and to establish its probability of predicting each user state correctly. The external parameters of this DTMC reflect the accuracy of the two ML models (lines 3--6), e.g., \textsf{p1\_ok\_correct} represents the probability that model M1 makes a correct prediction when the true user state is \textsf{ok}. The dependency parameters \texttt{pModel1} and \texttt{pModel2} specify the probabilities of using different ML models to monitor the user state. The ML model~1 is used with probability \texttt{pModel1} (by setting \texttt{m=1} in line~19, and therefore moving to the commands from lines~22--26), ML model~2 with probability \texttt{pModel2} (by setting \texttt{m=2} in line~19, and therefore moving to the commands starting in line~28), and both ML models are used as follows by \texttt{m=3} with probability $1-\texttt{pModel1}-\texttt{pModel2}$ in line~20: first model~1 generates a prediction, which serves as the output of the monitor if its verification by the verifier yields \texttt{true}, or otherwise model~2 is additionally used and its prediction becomes the output of the monitor.

\begin{lstlisting}[float, belowskip=-1.8\baselineskip, language={Prism}, numbers=left, rulesepcolor=\color{black}, rulecolor=\color{black}, breaklines=true, breakatwhitespace=true, firstnumber=1, firstline=1, 
caption={DTMC model of user monitoring component},
label={lst:RAD-user-perception-model}]
dtmc

// parameters of ML model 1/model 2/model 1 with verification
const double p1_ok_correct;
const double p1_notok_correct;
...
// Probabilities of using ML model 1 or 2
const double pModel1;
const double pModel2;

module UserPerception
  s : [0..4] init 0;      // step
  m : [1..3];             // operation mode
  ok : bool;              // true user state
  predOk : bool;          // predicted user state
  verified : bool;        // verifier result

  [] s=0 -> 0.5:(s'=1)&(ok'=true) + 0.5:(s'=1)&(ok'=false); 
  [] s=1 -> pModel1:(s'=2)&(m'=1) + pModel2:(s'=2)&(m'=2) +
            (1-pModel1-pModel2):(s'=2)&(m'=3);

  // Operation mode 1: use only ML model 1 
  [] s=2 & ok & m=1 -> p1_ok_correct:(s'=4)&(predOk'=true)
                  + (1-p1_ok_correct):(s'=4)&(predOk'=false);
  [] s=2 & !ok & m=1 -> p1_notok_correct:(s'=4)&(predOk'=false) 
                  + (1-p1_notok_correct):(s'=4)&(predOk'=true);

  // Operation mode 2: use only ML model 2 
  ...
  // Operation mode 3: use both ML models
  ... 

  [done] s=4 -> true;
endmodule

label "done" = s=4;
\end{lstlisting}

PMC applied to this DTMC and the PCTL-encoded properties below can be used to establish the probabilities that the user state is correctly predicted when the user's true state is \textsf{ok} and \texttt{notok}:
\begin{equation}
 \label{eq:RAD-vars}
\begin{array}{c}
  \textsf{pOkCorrect} = %\frac{
  P_{=?}[F\; \textsf{"done"} \wedge \textsf{ok} \wedge \textsf{predictedOk}]
  %}{P_{=?}[F\; \textsf{"done"} \wedge \textsf{ok}]} 
  \\[1mm]
  \textsf{pNotOkCorrect} = %\frac{
  P_{=?}[F\; \textsf{"done"} \wedge \neg\textsf{ok} \wedge \neg\textsf{predictedOk}]
  %}{ P_{=?}[F\; \textsf{"done"} \wedge \neg\textsf{ok}]}
  \end{array}
\end{equation}

%\smallskip
\noindent
\textbf{User monitor controller.} As shown by its DTMC model from Listing~\ref{lst:RAD-user-monitor-controller-model}, this RAD component uses the probabilities~\eqref{eq:RAD-vars} with which the user monitor outputs a true positive/negative, first to calculate the monitor's F1 score metric. and then to determine the probabilities with which the monitor should use its ML models~1 and~2:
\begin{equation}
    \label{eq:RAD-controller}
    \textsf{pModel1} =  P_{=?}[F\; s=1],\; \textsf{pModel2} =  P_{=?}[F\; s=2]
\end{equation}

\begin{lstlisting}[float, belowskip=-1.8\baselineskip, language={Prism}, numbers=left, rulesepcolor=\color{black}, rulecolor=\color{black}, breaklines=true, breakatwhitespace=true, firstnumber=1, firstline=1,
caption={DTMC model of user monitor controller},
label={lst:RAD-user-monitor-controller-model}]
dtmc

const double pOkCorrect;
const double pNotOkCorrect;
const double precision = 
                pNotOkCorrect/(0.5+pNotOkCorrect-pOkCorrect);
const double recall = 2*pNotOkCorrect;
const double F1 = 2*precision*recall/(precision+recall);

module ModelSelector
  s : [0..5] init 0;
  [] s=0 -> 0.47*F1:(s'=1)+(1-0.82*F1):(s'=2)+0.35*F1:(s'=3);
  [] s>0 -> true;
endmodule
\end{lstlisting}

%\smallskip
\noindent
\textbf{Dressing workflow.} The execution of a dressing session is modeled as a POMDP (Listing~\ref{lst:RAD-workflow-model}) because the true status of the user (\textsf{ok=true} or \textsf{ok=false}) is not observable. As such, \textsf{oK} does not appear in the list of observable variables from line~10.
The dressing starts with the garment picking by the robot (\texttt{s=1} in line~17). This succeeds with probability \texttt{pPickGarment} given by~\eqref{eq:RAD-pPickGarment}, and is followed by observing the user state, with the (unobservable) true user state \textsf{ok} and the predicted user state \textsf{predOk} appropriately set depending on the probability \textsf{pOk} that the user is in an \texttt{ok} state, and of the probabilities \eqref{eq:RAD-vars} of correctly predicting this user state (lines 18--21). Next, the system must decide between two actions (lines 23--31):
\squishlisttwo
\item \textsf{dressSlow}, which performs the user dressing slowly, taking longer but having a better chance of success if the user is \textsf{notok}; 
\item \textsf{dressFast}, which performs the dressing fast, but with a higher failure probability if the user is not \textsf{ok}.
\squishend
Depending on the selected action (i.e., on the POMDP \emph{policy}) and the user state, the dressing may succeed---in which case the success state in line~34 is reached, or not---in which case a retry is possible if allowed by the user (line~33). The workflow fails (line~35) if either the robot cannot pick the garment (line~17) or the user disagrees to dressing being retried after an unsuccessful attempt (line~33). 

%\begin{figure}
%    \centering
%\fbox{\includegraphics[width=\linewidth]{figures/sspomdp.pdf}}
%    \caption{Dressing workflow modelled as a POMDP}
%    \label{fig:RAD-workflow-model}
%\end{figure}%1.22-24 left aligned,line 31, remove spaces around + env->external 3.step->s 4. gap between the frame and the figure.

We note that the parameters \textsf{pPickGarment}, \textsf{pOkCorrect} and \textsf{pNotOkCorrect} of the POMDP take values obtained through the PMC of two other stochastic models presented in this section, reflecting the dependency of the POMDP on these models. All other POMDP parameters are external parameters. Finally, the choice between a \textsf{dressSlow}/\textsf{dressFast} action (i.e., the synthesis of a POMDP policy) is made by optimising a PCTL-encoded objective such as
\begin{equation}
    \label{eq:RAD-min-fail-property}
    \textsf{pFailMin} = Pmin_{=?} [F\; \textsf{step=6}],
\end{equation}
which requests the minimisation of the workflow failure probability.

\begin{lstlisting}[float, belowskip=-1.8\baselineskip, language={Prism}, numbers=left, rulesepcolor=\color{black}, rulecolor=\color{black}, breaklines=true, breakatwhitespace=true, firstnumber=1, firstline=1, linewidth=\hsize, xleftmargin=9pt, caption={Dressing workflow modelled as a POMDP},
label={lst:RAD-workflow-model}]
pomdp

const double pPickGarnment; // dependency parameter eq. (1)
const double pOkCorrect;    // dependency parameter eq. (2)
const double pNotOkCorrect; // dependency parameter eq. (2)
const double pOk;           // probability that user is ok
const double pDressOkSlow;  // slow-dress success prob/user ok
const double pDressOkFast;  // fast-dress success prob/user ok
...
observables s, predOk endobservables

module Workflow
  s : [1..6] init 1;       // workflow step
  ok : bool init true;     // true user state (hidden)
  predOk : bool init true; // predicted user state
  
  [pick] s=1 -> pPickGarnment:(s'=2) + (1-pPickGarnment):(s'=6);
  [obsv] s=2 -> pOk*2*pOkCorrect:(s'=3) +
     pOk*(1-2*pOkCorrect):(s'=3)&(predOk'=false)+
     (1-pOk)*2*pNotOkCorrect:(s'=3)&(ok'=false)&(predOk'=false)+
     (1-pOk)*(1-2*pNotOkCorrect):(s'=3)&(ok'=false);

  // dress slowly (higher success prob., longer time) or fast
  [dressSlow] s=3 & ok -> pDressOkSlow:(s'=5) + 
                          (1-pDressOkSlow):(s'=4);
  [dressFast] s=3 & ok -> pDressOkFast:(s'=5) + 
                          (1-pDressOkFast):(s'=4);
  [dressSlow] s=3 & !ok -> pDressNotOkSlow:(s'=5) + 
                           (1-pDressNotOkSlow):(s'=4);
  [dressFast] s=3 & !ok -> pDressNotOkFast:(s'=5) + 
                           (1-pDressNotOkFast):(s'=4);

  [retry] s=4 -> pRetryAllowed:(s'=3)+(1-pRetryAllowed):(s'=6);
  [succ] s=5 -> true;
  [fail] s=6 -> true;      
endmodule
\end{lstlisting}

\section{Theoretical foundation} \label{sec:foundation}

\subsection{Problem definition}

Our ULTIMATE framework supports the joint verification of sets of heterogeneous, interdependent stochastic models, such as the RAD models from Section~\ref{sec:motivating}. We refer to these as \emph{multi-model stochastic systems}, and provide a definition below.

\begin{definition}\label{def:ultimate}
A multi-model stochastic system is a tuple 
\begin{equation}
  U=(M,D,E), \label{eq:ULTIMATE-world-model}
\end{equation}
where: 
\squishlist
\item $M=\{m_1, m_2, \ldots, m_n\}$ is a set of $n>1$ stochastic models such that the transition probabilities/rates and rewards of each model $m_i\in M$ are defined by rational functions over a set $D_i=\{d^i_1,d^i_2,\ldots,d^i_{k_i}\}$ of $k_i\geq 0$ \emph{dependency parameters} and a set $E_i=\{e^i_1,e^i_2,\ldots,e^i_{l_i}\}$ of $l_i\geq 0$ \emph{external parameters};
\item $D=\{(m_i,d,m_j,\phi) \mid m_i,m_j\in M \wedge d\in D_i \wedge %\{d^i_1,d^i_2,\ldots,d^i_{k_i}\}, 
d=\mathit{pmc}(m_j,\phi) \}$ represents the set of \emph{dependency relationships} between the $n$ models, with each element  $(m_i,d,m_j,\phi)\in D$ specifying that the value of the dependency parameter $d\in D_i$ of $m_i$ is to be obtained by applying PMC to the property $\phi$ of model $m_j$;
\item $E=\{(m_i,e,f,O) \mid m_i\in M \wedge e\in E_i \wedge e=f(O)\}$ represents the set of \emph{external-parameter inferences}, with each element $(m_i,e,f,O)\in E$ specifying that the value of the external parameter $e\in E_i$ of $m_i$ is to be obtained by applying the (frequentist or Bayesian) inference function $f$ to the \emph{observations} (i.e., log entries or runtime measurements) $O$.
\squishend
\end{definition}

\begin{example}
\label{ex:RAD-multi-model}
The multi-model stochastic system for the RAD cyber-physical system from Section~\ref{sec:motivating} is 
$
  U_\mathit{RAD} =(M,D,E), 
$
where:
\squishlist
\item $M=\{\mathit{m}_\mathit{gp}, \mathit{m}_\mathit{um}, \mathit{m}_\mathit{umc}, \mathit{m}_\mathit{dp}\}$ is the set of stochastic models comprising the garment picking CMTC in Listing~\ref{lst:RAD-garment-picking-model} ($\mathit{m}_\mathit{gp}$), the user monitoring DTMC in Listing~\ref{lst:RAD-user-perception-model} ($\mathit{m}_\mathit{um}$), the user monitor controller DTMC in Listing~\ref{lst:RAD-user-monitor-controller-model} ($\mathit{m}_\mathit{umc}$), and the dressing workflow POMDP  model from Listing~\ref{lst:RAD-workflow-model} ($\mathit{m}_\mathit{dp}$);
\item $D$ contains elements that define the 7~dependency relationships from Figure~\ref{fig:RAD-models}; for example, $(\mathit{m}_\mathit{dp},\textsf{pPickGarment},\mathit{m}_\mathit{gp},$ \mbox{$P_{=?}[F^{[0,90]}\; \textsf{"success"}])\in D$ specifies that} the dependency parameter \textsf{pPickGarment} of $\mathit{m}_\mathit{dp}$ needs to be calculated by applying PMC to property~\eqref{eq:RAD-pPickGarment} of $\mathit{m}_\mathit{gp}$; 
\item $E$ contains elements that specify the inference functions and observations to be used for estimating the external parameters of the RAD models, e.g., $(\mathit{m}_\mathit{gp},\textsf{rPick},\mathit{meanRate},\{47,92,$ $61,\ldots\})$ to specify that the external parameter \textsf{rPick} of $\mathit{m}_\mathit{gp}$ should be approximated as the mean rate of a process with the observed durations from the set $\{47,92,61,\ldots\}$. Other options for determining the value of an external parameter include using a Bayesian estimator to derive a posterior value from a prior value and a set of observations, or simply using a predefined value.
\squishend
\end{example}

Given a multi-model stochastic system $U=(M,D,E)$ and a formally specified property $\phi_v$ of a model $m_v\in M$, the verification problem tackled by our ULTIMATE framework is to establish the value of property $\phi_v$, i.e., to compute $\mathit{pmc}(m_v,\phi_v)$. Solving this verification problem is challenging as it may require: (i)~the PMC of additional models from $M$ to establish the value of the dependency parameters of model $m_v$, which (as detailed in Section~\ref{subsect:algorithm}) is particularly complex in the presence of circular model interdependencies, (ii)~the synthesis of suitable policies for any verified models (e.g., MPDs and POMDPs) that contain nondeterminism, and (iii)~the estimation of the external parameters of all models involved (directly or through model interdependencies) in the verification.

\subsection{ULTIMATE verification algorithm} \label{subsect:algorithm}

Before presenting our algorithm for the verification of ULTIMATE multi-model stochastic systems, we need to define two concepts.

\begin{definition}\label{def:graph1}
The \emph{dependency graph} induced by an ULTIMATE multi-model stochastic system $U=(M,D,E)$ is the directed graph $\mathcal{G}\!=\!(\mathcal{V},\mathcal{E})$ with vertex set $\mathcal{V}\!=\!\{1,2,\ldots,n\}$ comprising a vertex for each model $m\!\in\! M$, 
and edge set $\mathcal{E}=\{(i,j) \mid  \exists (m_i,d,m_j,\phi)\!\in\! D\}$.
\end{definition}

\begin{definition}\label{def:graph2}
Given an ULTIMATE multi-model stochastic system $U=(M,D,E)$ and its induced dependency graph $\mathcal{G}=(\mathcal{V},\mathcal{E})$, the \emph{strongly connected component} associated with model $m_i\in M$ is the subset of models $\mathit{SCC}(m_i)=\{m_j \mid m_j\in M \wedge j\in \mathcal{V}_i\}$, where $\mathcal{V}_i\subseteq \mathcal{V}$ is the set of vertices from the strongly connected component of $G$ that includes vertex $i$. 
\end{definition}

\begin{example}
\label{ex:dep-graph-and-SCC}
The vertices and edges of the dependency graph $\mathcal{G}=(\mathcal{V},\mathcal{E})$ induced by the RAD multi-model stochastic system $U_\mathit{RAD}$ from Example~\ref{ex:RAD-multi-model} are $\mathcal{V}=\{1,2,3,4\}$ (corresponding to the four $U_\mathit{RAD}$ models) and $\mathcal{E}=\{(1,4),(2,3),(2,4),(3,2)\}$ (corresponding to the arrows from Figure~\ref{fig:RAD-models}). Accordingly, the strongly connected components associated with the four models are $\mathit{SCC}(\mathit{m}_\mathit{gp})=\{\mathit{m}_\mathit{gp}\}$, $\mathit{SCC}(\mathit{m}_\mathit{um})=\mathit{SCC}(\mathit{m}_\mathit{umc})=\{\mathit{m}_\mathit{um},\mathit{m}_\mathit{umc}\}$ and $\mathit{SCC}(\mathit{m}_\mathit{dp})=\{\mathit{m}_\mathit{dp}\}$.
\end{example}

\bigskip\noindent
We note that the set of strongly connected components 
\begin{equation}
  \label{eq:scc}
  \mathit{SCC}=\{\mathit{SCC}(m_1),\mathit{SCC}(m_2),\ldots,\mathit{SCC}(m_n)\}
\end{equation}  
associated with all models of a multi-model stochastic system $U=(M,D,E)$ can be computed efficiently in linear time. This computation involves first assembling the dependency graph $\mathcal{G}=(\mathcal{V},\mathcal{E})$ of $U$ (in linear, $\mathrm{O}(\#M+\#D)$ time, by examining each model from $M$ and each dependency from $D$ once), and then applying Tarjan's strongly connected component detection algorithm~\cite{Tarjan1972DepthFirstSA} (which takes linear, $\mathrm{O}(\#V+\#E)$ time) to this graph.

\begin{algorithm}
\caption{Verification of multi-model stochastic systems }\label{algo:verification}
\begin{algorithmic}[1]
\Function{\textsc{Verify}}{$(M,D,E), m_v, \phi_v, \mathit{SCC}$}
   \For{$m\in\mathit{SCC}(m_v)$} \label{l:verify-outer-for-start}
       \For{$(m,d,m',\phi)\in D$} \label{l:verify-inner-for1-start}
          \If{$m'\notin\mathit{SCC}(m_v)$} \label{l:verify-if-start}
             \State \textsc{SetParam}($m,d,\textsc{Verify}((M,D,R),m',\phi,\mathit{SCC})$  \label{l:set-param1}
          \EndIf \label{l:verify-if-end}
       \EndFor \label{l:verify-inner-for1-end}
       \For{$(m,e,f,O)\in E$} \label{l:verify-inner-for2-start}
            \State \textsc{SetParam}($m,e,f(O)$) \label{l:verify-set-external-param}
       \EndFor \label{l:verify-inner-for2-end}
    \EndFor \label{l:verify-outer-for-end}
    \If{$\mathit{SCC}(m_v)\neq\{m_v\}$} \label{l:verify-outer-if-start}
       \State \textsc{VerifySCC}($(M,D,E),\mathit{SCC}(m_v)$) \label{l:VerifySCC-call}
    \EndIf \label{l:verify-outer-if-end}
    \State \Return \textsc{PMC}($m_v, \phi_v$) \label{l:verif-return}
\EndFunction
\State
\Function{\textsc{VerifySCC}}{$(M,D,E), \mathit{M'}$}
   \State $\mathit{Equations}\gets\{\}$, $\mathit{ParamList}\gets\{\}$ \label{l:VerifySCC-start}
   \For{$m\in M'$} \label{l:verify-scc-outer-for-start}
      \For{$(m,d,m',\phi)\in D$} \label{l:verify-scc-inner-for-start}
          \If{$m'\in M'$} \label{l:verify-scc-check}
             \If{$\textsc{ParametricVerifFeasible}(M',D)$} 
               \State $\mathit{equation}\gets \textrm{`}d=\textsc{ParametricMC}(m',\phi)\textrm{'}$ \label{l:verif-scc-eq1}
             \Else
                \State $\mathit{equation}\gets \textrm{`}d=\mathit{pmc}(m',\phi)\textrm{'}$ \label{l:verif-scc-eq2}
             \EndIf
             \State $\mathit{Equations}\gets\mathit{Equations}\cup \{\mathit{equation}\}$
             \State $\mathit{ParamList}\gets\mathit{ParamList}\cup\{(m,d)\}$
          \EndIf
       \EndFor
    \EndFor \label{l:verify-scc-outer-for-end}
    \State $\mathit{Solution} \gets \textsc{Solve}(\mathit{Equations})$ \label{l:verify-scc-solution}
    \For{$(m,d)\in\mathit{ParamList}$} \label{l:verif-scc-final-for-start}
       \State \textsc{SetParam}($m,d,\mathit{Solution}(d)$) 
     \EndFor \label{l:verif-scc-final-for-end}
\EndFunction
\end{algorithmic}
\end{algorithm}

Having defined these concepts, we can now present our ULTIMATE verification process, which is carried out by function \textsc{Verify} from Algorithm~\ref{algo:verification}. This function takes four arguments---the multi-model stochastic system $U=(M,D,E)$, model $m_v\in M$ and property $\phi_v$ to be verified, and $U$'s set $\mathit{SCC}$ of strongly connected components~\eqref{eq:scc}---and performs the verification of property $\phi_v$ of model $m_v$ in the three stages described below.

\smallskip\noindent
\textbf{Stage~1:} In this stage, the for loop in lines~\ref{l:verify-outer-for-start}--\ref{l:verify-outer-for-end} iterates through each model $m$ from the strongly connect component $\mathit{SCC}(m_v)$ associated with the model under verification, using:
\squishlisttwo
\item the inner for loop in lines~\ref{l:verify-inner-for1-start}--\ref{l:verify-inner-for1-end} to extract each dependency parameter $d$ of $m$ (line~\ref{l:verify-inner-for1-start}) and, if the origin of the dependency is a model $m'$ external to the strongly connected component (line~\ref{l:verify-if-start}), to obtain the value of that parameter by invoking \textsc{Verify} recursively, and to set the parameter $d$ to that value (line~\ref{l:set-param1});
\item the inner for loop in lines~\ref{l:verify-inner-for2-start}--\ref{l:verify-inner-for2-end} to extract each external parameter $e$ of $m$ (line~\ref{l:verify-inner-for2-start}), and to obtain and set the value of that parameter according to its specification from Definition~\ref{def:ultimate}.
\squishend

\smallskip\noindent
\textbf{Stage~2:} In this stage, the if statement in lines~\ref{l:verify-outer-if-start}--\ref{l:verify-outer-if-end} resolves any co-dependencies among the models of the strongly connected component $\mathit{SCC}(m_v)$ associated with the model under verification. To that end, the function \textsc{VerifySCC} is invoked in line~\ref{l:VerifySCC-call} if $\mathit{SCC}(m_v)$ is not a ``degenerate'' strongly connected component containing only model $m_v$. Given a strongly connected component $M'$ comprising models whose dependency parameters depend on models from outside $M'$ and external parameters are already computed, lines~\ref{l:VerifySCC-start}--\ref{l:verify-scc-outer-for-end} of this function assemble a system of equations whose solution in line~\ref{l:verify-scc-solution} provides the values for setting the remaining dependency parameters of the models from $M'$ in lines~\ref{l:verif-scc-final-for-start}--\ref{l:verif-scc-final-for-end}. The system of $\mathit{Equations}$ solved in line~\ref{l:verify-scc-solution} is assembled by iterating through every model from $M'$ (line~\ref{l:verify-scc-outer-for-start}) and dependency parameter of that model (line~\ref{l:verify-scc-inner-for-start}), and adding a new $\mathit{equation}$ to $\mathit{Equations}$ if the origin of the dependency is a model from the strongly connected component $M'$ (line~\ref{l:verify-scc-check}). This $\mathit{equation}$ can take one of two forms. If \emph{all} dependency parameters between pairs of models from $M'$ (i.e., all parameters for which an equation needs to be assembled) require the verification of a property that can be analysed using parametric model checking, then the auxiliary function \textsc{ParametricVerifFeasible} returns \textsf{true},\footnote{At the time of writing, parametric model checking can handle DTMC properties without inner probabilistic operators, and non-transient CTMC properties, so this is what the auxiliary function \textsc{ParametricVerifFeasible} (which we do not include due to space constraints) needs to check.} and parametric model checking is used to obtain an algebraic $\mathit{equation}$ in line~\ref{l:verif-scc-eq1}. Otherwise, line~\ref{l:verif-scc-eq2} creates an $\mathit{equation}$ whose right-hand side (i.e., $\mathit{pmc}(m',\phi)$) can only be evaluated by applying probabilistic model checking to instances of $m'$ whose unknown dependency parameters are set to specific combinations of values. 
%
% OLD TEXT (Radu)
% We note that the \textsc{Solve} function from line~\ref{l:verify-scc-solution} must handle the $\mathit{Equations}$ passed to it by using numeric solvers and optimisers (see Figure~\ref{fig:approach}) selected according to whether these $\mathit{Equations}$ are of the first or the second form, and that the selection of the valid solution (by the engineer performing the verification) is needed when multiple solutions are mathematically feasible. 
%

Depending on whether the $\mathit{Equations}$ are of the first or the second form, the \textsc{Solve} function from line~\ref{l:verify-scc-solution} employs numeric solvers and optimisers as appropriate (see Figure~\ref{fig:approach}). 
In particular, when these $\mathit{Equations}$ contain algebraic formulae, \textsc{Solve} resorts to solving a system of nonlinear equations either analytically (e.g., via algebraic manipulation) or numerically (e.g., using the Newton-Raphson optimisation method~\cite{nocedal1999numerical}). 
When \textsc{ParametricVerifFeasible} returns \textsf{false} and probabilistic model checking is used, the \textsc{Solve} function leverages derivative-free optimization (e.g., Powell’s optimization method~\cite{powell2007view}) to estimate the dependency parameter values by minimising the sum of squared residuals between those values and the outcome of $\mathit{pmc}(m',\phi)$.
%We also note that when multiple solutions are mathematically feasible, the selection of a valid solution by the engineer performing the verification is needed.

\smallskip\noindent
\textbf{Stage~3:} Finally, with all dependency and external parameters of model $m_v$ resolved, the final stage of \textsc{Verify} invokes a probabilistic model checking engine to obtain and return the required verification result in line~\ref{l:verif-return}.

\subsection{Correctness of the verification algorithm}

To demonstrate the correctness of Algorithm~\ref{algo:verification}, we need the following definition.

\begin{definition}
\label{def:dep-level}
Consider a multi-model stochastic system $U=(M,D,$ $E)$, and the directed acyclic graph (i.e., the dag) comprising a vertex for each strongly connected component of $U$ and an edge between each pair of vertices $(v,v')$ for which the strongly connected component associated with $v'$ contains a model with a dependency parameter defined in terms of a property of a model from the strongly connected component associated with $v$. The \emph{dependency level} of a model $m\in M$ is defined as the length of the longest path from a vertex of this dag to the vertex associated with $\mathit{SCC}(m)$, the strongly connected component of $m$.
\end{definition}

\begin{example}
    Consider the multi-model stochastic system $U_\mathit{RAD}$ from Example~\ref{ex:RAD-multi-model}. As discussed in Example~\ref{ex:dep-graph-and-SCC}, $U_\mathit{RAD}$ has three strongly connected components: $\{\mathit{m}_\mathit{gp}\}$, $\{\mathit{m}_\mathit{um},\mathit{m}_\mathit{umc}\}$ and $\{\mathit{m}_\mathit{dp}\}$. As such, its dag from Definition~\ref{def:dep-level} will have three vertices, $v_1$, $v_2$ and $v_3$, associated (in order) with these strongly connected components, and (given the model dependencies depicted in Figure~\ref{fig:RAD-models}) the edges $(v_1,v_3)$ and $(v_2,v_3)$. Models $\mathit{m}_\mathit{gp}$, $\mathit{m}_\mathit{um}$ and $\mathit{m}_\mathit{umc}$ belong to strongly connected components associated with vertices $v_1$ and $v_2$, for which the maximum path length from another vertex in the dag is $0$. Therefore, their dependency level is $0$. In contrast, model $\mathit{m}_\mathit{dp}$ belongs to the strongly connected component associated with vertex $v_3$, for which the maximum path length from another vertex in the dag is $1$, and thus the dependency level of this model is $1$.
\end{example}

\begin{theorem}
    Function \textsc{Verify} from Algorithm~\ref{algo:verification} terminates and returns the correct value of property $\phi_v$ of stochastic model $m_v$.
\end{theorem}

\vspace*{-4mm}
\begin{proof}
    We prove this result by induction over the dependency level of $m_v$. For the \textbf{base step}, which corresponds to $m_v$ having dependency level $0$, we need to consider two cases. 
    
    The \textbf{first base case} is when $m_v$ belongs to a `degenerate'' strongly connected component, i.e., $\mathit{SCC}(m_v)=\{m_v\}$. The garment picking CTMC from Figure~\ref{fig:RAD-models} is an example of such a model. In this case, the for loop in lines~\ref{l:verify-outer-for-start}--\ref{l:verify-outer-for-end} is only executed once, for $m=m_v$. This execution skips the inner for loop from lines~\ref{l:verify-inner-for1-start}--\ref{l:verify-inner-for1-end} (since $m_v$ has no dependency parameter), and sets all the external parameters of $m_v$ in the for loop from lines~\ref{l:verify-inner-for2-start}--\ref{l:verify-inner-for2-end}. The function then moves directly to computing the property $\phi_v$ of model $m_v$ in line~\ref{l:verif-return}, since the if statement condition from line~\ref{l:verify-outer-if-start} does not hold. As $m_v$ has no dependency parameters and all its external parameters were already set, this computation is feasible, and its result---the correct value of $\phi_v$---is obtained and returned, ending the execution of \textsc{Verify}. 
    
    The \textbf{second base case} is when $m_v$ belongs to a strongly connected component that also contains additional models, i.e., $\mathit{SCC}(m_v)\setminus\{m_v\}\neq\emptyset$. The user monitoring DTMC from Figure~\ref{fig:RAD-models} is an example of such a model. In this case, the outer for loop in lines~\ref{l:verify-outer-for-start}--\ref{l:verify-outer-for-end} iterates through every model $m\in \mathit{SCC}(m_v)$. Each such iteration: (i)~skips the recursive invocation of \textsc{Verify} from line~\ref{l:set-param1} because the origin model $m'$ of every dependency parameter is in $\mathit{SCC}(m_v)$, so the condition of the if statement from line~\ref{l:verify-if-start} is \textsf{false}; and (ii)~calculates and sets all the external parameters of $m$ in the for loop from lines~\ref{l:verify-inner-for2-start}--\ref{l:verify-inner-for2-end}. With all these parameters resolved, \textsc{VerifySCC} is called in line~\ref{l:VerifySCC-call}, as $\mathit{SCC}(m_v)\neq\{m_v\}$. As a result, each dependency parameter of these models is visited precisely once by the two nested for loops from lines~\ref{l:verify-scc-outer-for-start}--\ref{l:verify-scc-outer-for-end}, so that a system of equations comprising as many equations as there are dependency parameters is assembled. The relevant solution to these equations (obtained in line~\ref{l:verify-scc-solution}) is used to set the values of all dependency parameters of all the models from $\mathit{SCC}(m_v)$ in lines~\ref{l:verif-scc-final-for-start}--\ref{l:verif-scc-final-for-end}, after which \textsc{VerifySCC} exits, enabling the \textsc{Verify} function to then obtain and return the value of property $m_v$ in line~\ref{l:verif-return}, and to terminate. 

    For the \textbf{inductive step}, we assume that the theorem holds for all verified models with dependency level up to $N\geq 0$, and we consider the verification of a model $m_v$ with dependency level $N+1$. In this case, every dependency parameter $d$ of $m_v$ and of any other model from $\mathit{SCC}(m_v)$ whose origin is a model $m'$ from outside $\mathit{SCC}(m_v)$ is visited once by the two nested for loops starting in lines~\ref{l:verify-outer-for-start} and~\ref{l:verify-inner-for1-start}, and is   
    computed through a recursive \textsc{Verify} invocation in line~\ref{l:set-param1}. Since the dependency level of model $m'$ is smaller than that of $m_v$, and thus between $0$ and $N$, our assumption implies that this recursive invocation of \textsc{Verify} terminates and returns the correct verification result required to set the dependency parameter $d$ in line~\ref{l:set-param1}. Hence, all dependency parameters of models from $\mathit{SCC}(m_v)$ that depend on models from outside $\mathit{SCC}(m_v)$ are correctly set by the for loop in lines~\ref{l:verify-inner-for1-start}--\ref{l:verify-inner-for1-end}, all external parameters of these models are calculated and set by the for loop from lines~\ref{l:verify-inner-for2-start}--\ref{l:verify-inner-for2-end}, and (for the same reasons as in the second base case) the dependency parameters that depend on models from within $\mathit{SCC}(m_v)$ are computed and set, if needed because $\mathit{SCC}(m_v)\neq\{m_v\}$, by the invocation of \textsc{VerifySCC} from line~\ref{l:VerifySCC-call}. As such, the verification of property $\phi_v$ of $m_v$ can be performed successfully in line~\ref{l:verif-return}, and the algorithm terminates---which completes the proof of the inductive step.
\end{proof}

We end this section by noting that a formal complexity analysis of our verification algorithm is not possible because of its use of multiple reasoning engines whose configuration can influence their execution time significantly. These include (Figure~\ref{fig:approach}) probabilistic and parametric model checkers (lines~\ref{l:verif-return} and~\ref{l:verif-scc-eq1}), numeric solvers and optimisers (line~\ref{l:verify-scc-solution}), and frequentist and Bayesian inference functions (line~\ref{l:verify-set-external-param}). To compensate for this, we provide an empirical assessment of the algorithm performance in Section~\ref{sec:evaluation}.

\section{Implementation \label{sec:implementation}}

\begin{figure}
    \centering
\includegraphics[width=\linewidth]{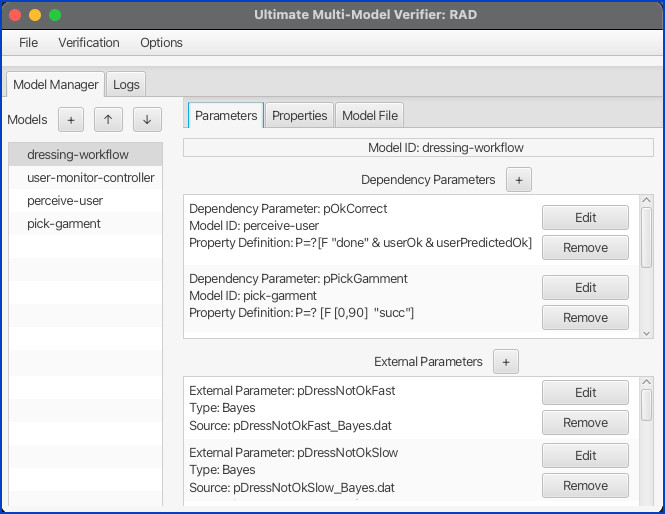}\\[-1.2mm]

$\downarrow$

\includegraphics[width=\linewidth]{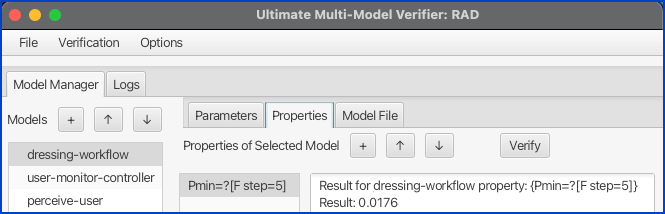}
    \caption{RAD system verification using the ULTIMATE tool}\label{fig:gui}

\vspace*{-2.5mm}
\end{figure}

To support the use and adoption of our verification framework, we developed an open-source ULTIMATE verification tool~\cite{ULTIMATE-repo-non-anonymised} with the architecture from Figure~\ref{fig:approach}. At the core of this tool is a Java implementation of the verification functions from Algorithm~\ref{algo:verification}, and a graphical user interface (Figure~\ref{fig:gui}) that enables users to: (i)~define ULTIMATE multi-model stochastic systems (with their component stochastic models, dependency parameters, and external parameters); (ii)~specify properties of these systems that require verification; (iii)~run verification sessions; and (iv)~examine both end and intermediate verification results. 

Our verification tool can handle multi-model stochastic systems comprising combinations of all the model types from Table~\ref{table:model-types}, with each dependency parameter defined in the rewards-extended probabilistic temporal logic(s) allowed by the type of the model whose analysis determines the parameter value (i.e., by the type of the model $m_j$ from Definition~\ref{def:ultimate}). To that end, the tool invokes as needed: the probabilistic and parametric model checkers PRISM~\cite{KNP11} and Storm~\cite{hensel2022probabilistic}, and the stochastic games model checker PRISM-games~\cite{Kwiatkowska18PW}; the numeric solvers and optimisers from the Scipy open-source optimisation library; and our Java implementations of the KAMI Bayesian estimator from~\cite{epifani2009model,10.1007/978-3-642-02351-4_5} and of simple (frequentist) mean functions.

\section{Case studies and experiments} \label{sec:evaluation}

We used the ULTIMATE framework and tool in five case studies (Table~\ref{table:case-studies}) covering complex systems from a broad range of application domains. Further details and supporting material (models, verified properties, external parameter datasets, descriptions) are available in our public online repository~\cite{ULTIMATE-repo-non-anonymised}. In this section, we summarise these case studies to demonstrate the benefits and versatility of our framework, and we assess its performance.

\begin{table*}
    \caption{Case study summary (see~\cite{ULTIMATE-repo-non-anonymised} for further details and supporting material)}
    \label{table:case-studies}
    \def\tabcolsep{2pt}
\centering
\sffamily
{\fontsize{8}{9.5}\selectfont 

\vspace*{-2mm}
\begin{tabular}{P{0.8cm} P{10.2cm} P{4.5cm} P{1.2cm}}
\toprule
\textbf{ID} & \textbf{Multi-model stochastic system} & \textbf{Sample verified $\textsf{model}:\textsf{property}$$^{\dagger}$} & \textbf{Time$^{\ddagger}$ (s)}\\ \midrule
RAD & $U_\mathit{RAD}$ multi-model from Example~\ref{ex:RAD-multi-model} & $\mathit{m}_\mathit{dp} : Pmin_{=?} [F\; \textsf{step=6}]$ & 3.10s \\ \midrule
SMD & $U_\mathit{SMD}=(M,D,E)$, where: & $m_\mathit{ms}:P_{=?}[F\; \textsf{largeObject}\wedge \textsf{detected}]/$ & 2.30s\\
& -- $M=\{m_\mathit{ms},m_\mathit{sl}\}$ & \hspace*{14mm}$P_{=?}[F\; \textsf{largeObject}])$\\
& -- $D=\{(m_\mathit{ms},\textsf{pLow},m_\mathit{sl},R^\textsf{low}_{=?} [C^{\leq 3600}]/3600),(m_\mathit{ms},\textsf{pMed},m_\mathit{sl},R^\textsf{med}_{=?} [C^{\leq 3600}]/3600)$, & $m_\mathit{sl}:R^\textsf{power}_{=?} [F\;\textsf{done}]$ & 3.00s\\
& \hspace*{8.5mm} $(m_\mathit{sl},\textsf{pDetect},m_\mathit{ms},P_{=?}[F\; \textsf{detected}])\}$ \\
& -- $E=\{(m_\mathit{ms},\textsf{pLargeObject},\mathit{predefined},0.05)\}$\\ \midrule
Robo- & $U_\mathit{RoboFleet}=(M,D,E)$, where: &
$m_\mathit{sup} : R^\mathsf{failures}_{=?}[F\; \mathsf{done}]$ & 0.03s\\
Fleet & -- $M=\{m_{r_1}, m_{r_2}, \ldots, m_{r_N}, m_\mathit{sup}\}$ &
$m_\mathit{sup} : R^\mathsf{cost}_{=?}[F\; \mathsf{done}]$ & 0.03s \\
& -- $D=\{(m_\mathit{sup},\textsf{pR1},m_{r_1},Pmax_{=?}[F\; \textsf{done}]),\ldots,(m_\mathit{sup},\textsf{pRN},m_{r_N},Pmax_{=?}[F\; \textsf{done}])\}$\\
& -- $E=\{(m_\mathit{sup},\mathsf{nAttempts},\mathit{predefined},3)\}$\\ \midrule
DPM- & $U_\mathit{DPM}=(M,D,E)$, where: & $R^\mathsf{time}_{=?}[F\; \mathsf{done}]$ & 1.50s\\
FX  & -- $M=\{m_\mathit{dpm}, m_\mathit{fx}\}$ \\
& -- $D=\{(m_\mathit{dpm},\mathsf{diskOps},m_\mathit{fx},R^\mathsf{ops}_{=?}[F\; \mathsf{done}]),(m_\mathit{fx},\mathsf{avrQueueDiskOps},m_\mathit{dpm},R^\mathsf{size}_{=?} [S])\}$ \\
& -- $E=\{(m_\mathit{dpm},\mathsf{pIdle2Sleep},\mathit{predefined},0.65), \dots%, (m_\mathit{dpm},\mathsf{service},\mathit{predefined},0.90)
\}$\\ \midrule
RoCo & $U_\mathit{RoCo}=(M,D,E)$, where: & $\langle \langle robot1:robot2\rangle\rangle \mathsf{max}_{=?}$  & 2.60s\\
& -- $M=\{m_\mathit{rob}, m_\mathit{od}\}$ & \hspace*{5mm}$(P[ \neg \mathsf{crash}\; U^{\leq k}\; \mathsf{goal1}] +$ \\
& -- $D=\{(m_\mathit{rob}, q,m_\mathit{od},P_{=?}[F\; \mathsf{obstacleAhead}])\}$ & \hspace*{10mm}$P[ \neg \mathsf{crash}\; U^{\leq k}\; \mathsf{goal2}])$ \\
& -- $E=\{(m_\mathit{od},\mathsf{pObstacle},\mathit{predefined},0.05)\}$ \\
\bottomrule
\multicolumn{4}{l}{$^{\dagger}$The reward formula $R_{=?}^\mathsf{rwd}[F\; \phi ]$ denotes the expected reward $\mathsf{rwd}$ cumulated to reach a future state of the verified model in which the formula $\phi$ holds}\\
\multicolumn{4}{l}{$^{\ddagger}$Total time to verify the sample property on a MacBook Pro with M1 Pro chip and 32GB RAM, averaged over 10 executions of Algorithm~\ref{algo:verification}}
\end{tabular}
}
\end{table*}

\begin{figure*}
    \centering
    \includegraphics[width=0.78\textwidth]{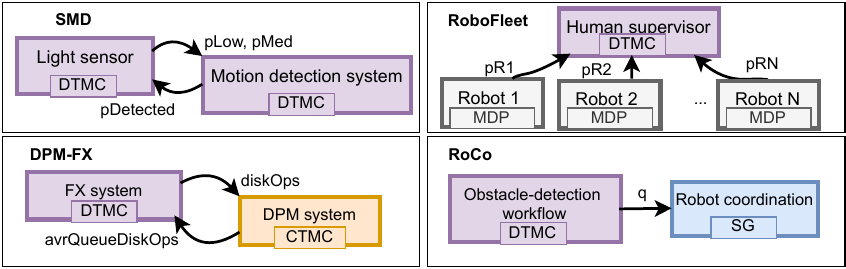}
    \caption{Stochastic models and dependencies for the SMD, RoboFleet, DPM-FX and RoCo case studies}\label{fig:case-studies}
\end{figure*}

\begin{figure*}
    \centering
\hspace*{9mm}\includegraphics[height=4cm]{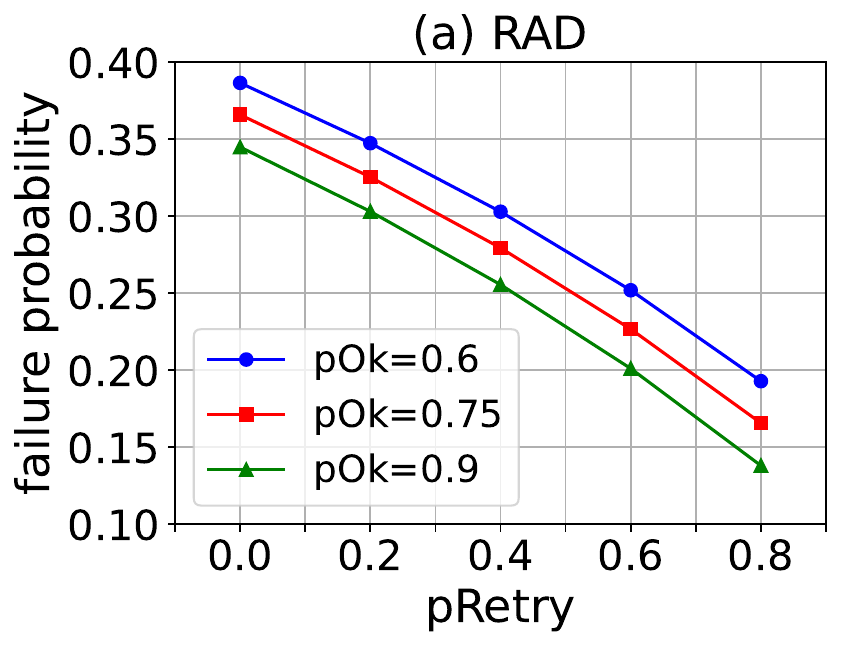}\hspace*{14mm}
\includegraphics[height=4cm]{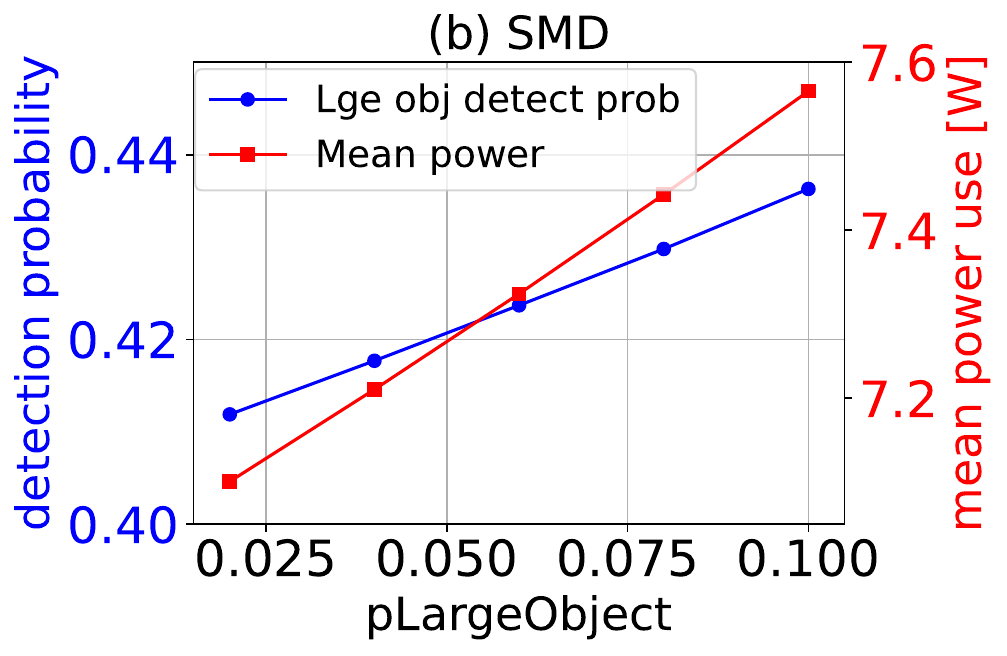}

\vspace*{2mm}
\includegraphics[height=4cm]{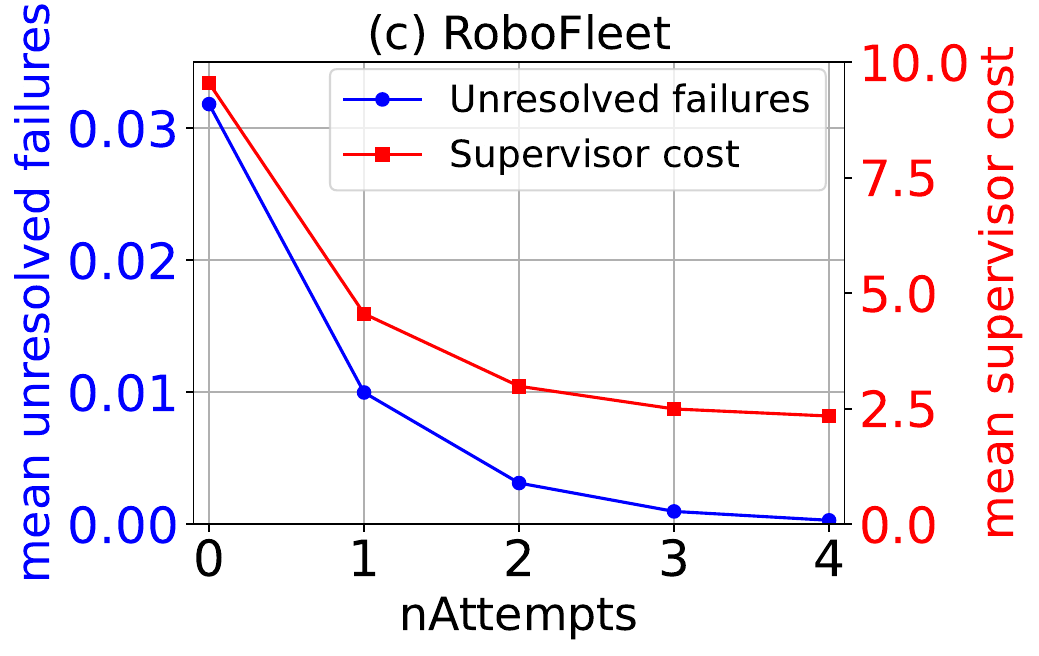}\hspace*{4mm}
\includegraphics[height=4cm]{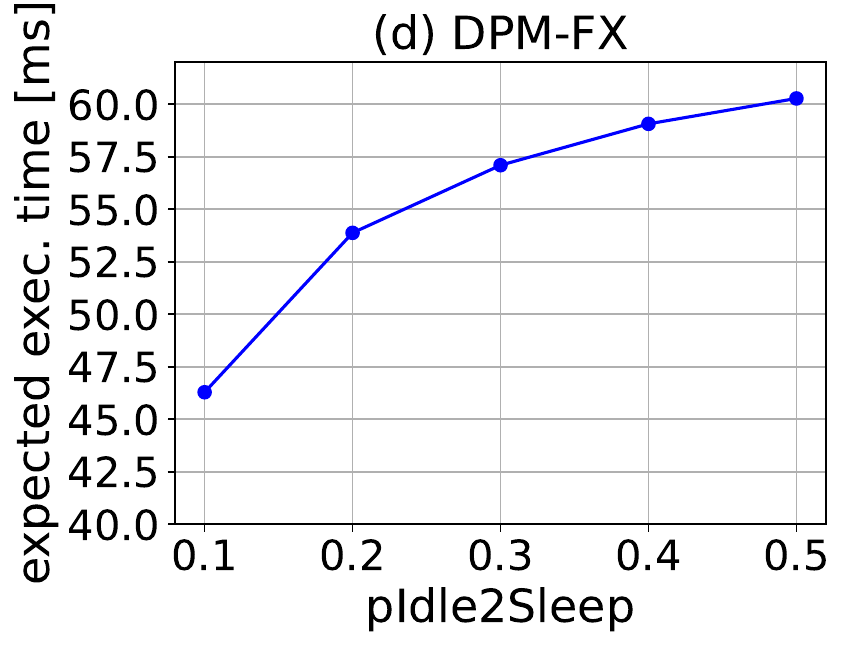}\hspace*{4mm}
\includegraphics[height=4cm]{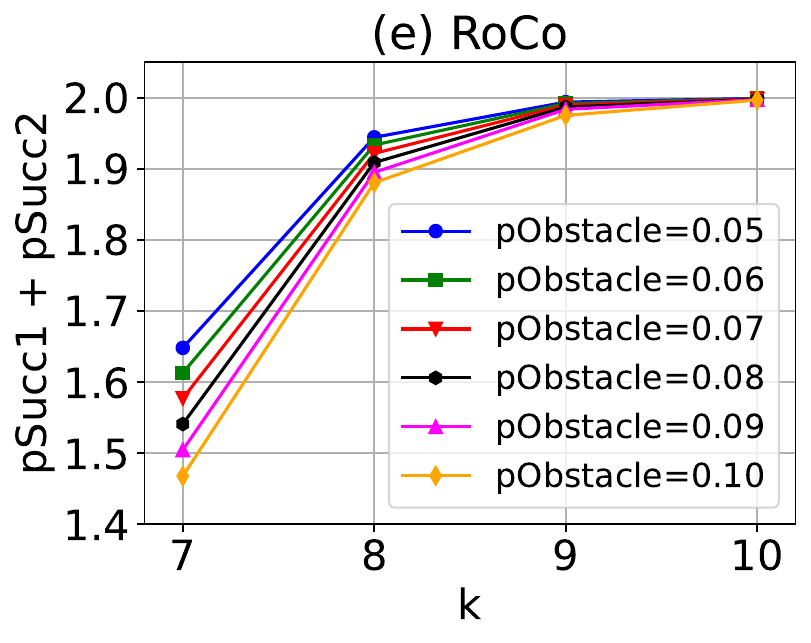}

\vspace*{-1.5mm}
\caption{Experimental results}
    \label{fig:results}
\end{figure*}

\medskip\noindent
\textbf{Robot assistive dressing (RAD).} This case study involved the verification of the RAD system from our motivating example introduced in Section~\ref{sec:motivating}. We verified the property~\eqref{eq:RAD-min-fail-property} of the dressing-process POMDP from Listing~\ref{lst:RAD-workflow-model} for different values of:
\squishlisttwo
\item the $\textit{pOk}$ probability that the user state is ``ok'' in this POMDP,
\item the garment-picking CTMC's $\textit{pRetry}$ probability (Listing~\ref{lst:RAD-garment-picking-model}),
\squishend
synthesising RAD controller policies that minimise the failure probability of the dressing process for these combinations of parameter values---see Figure~\ref{fig:results}a.

\medskip\noindent
\textbf{Smart motion detection (SMD).} Our second case study verifies a cyber-physical system comprising two co-dependent components: a night-time motion sensor, and a smart lighting component. The motion sensor is activated every few seconds, and triggers an alarm if it detects a moving object. This detection happens with a probability that depends on two factors. The first factor is the actual scenario encountered when the sensor is activated: (i)~large, relevant moving object (intruder, cat, dog, etc.) present in the monitored area, (ii)~small, irrelevant object (e.g., leaves moving due to wind) present, or (iii)~no moving object present. The second factor is the level of lighting in the monitored area. This can be low (with probability \textsf{pLow}), medium (with probability \textsf{pMed}), or high (with probability $\textsf{pHigh}=1-\textsf{pLow}-\textsf{pMed}$), where these three probabilities are determined by the operation of the smart lighting component. This component decides a (potentially new) level of lighting every half minute, based on (i)~the current level of lighting, and (ii)~the output of the motion sensor (i.e., moving object detected or not) when the decision is made. As such, the lighting level depends on the probability $\texttt{pDetect}$ that a moving object is detected by the motion sensor. With the behaviour of each component modelled as a DTMC ($m_\mathit{ms}$ for the motion sensor, and $m_\mathit{sl}$ for the smart lighting), we assembled the ULTIMATE multi-model stochastic system shown in Table~\ref{table:case-studies}, and used it to establish (Figure~\ref{fig:results}b) (i)~the probability that a large object is detected by the SMD system, and (ii)~the power consumption for the smart lighting component, for a range of probabilities that a large object is present in the monitored area.

\medskip\noindent
\textbf{Mobile robot fleet (RoboFleet).} In this case study, we consider a human-supervised fleet of $N$ mobile robots. Each robot performs an independent mission within a grid world in which it can move up, down, left or right between locations, subject to remaining within the bounds of the grid world, and to not entering locations that contain obstacles. The mission involves navigating from an initial location to a goal location where the robot needs to perform a task. At each location, the robot may become stuck with a small probability that depends of the distance between that location and any nearby obstacles. We use separate MDPs ($m_{r_1}$, $m_{r_2}$, \ldots, $m_{r_N}$) to model the robots and their missions, and to derive the maximum mission success probabilities \textsf{pR1}, \textsf{pR2}, \ldots, \textsf{pRN}, and the associated optimal MDP policies, which correspond to the sequences of movements that the robots should use to maximise their mission success. A human supervisor whose operation depends on these probabilities is monitoring the robot fleet, trying to unstuck blocked robots, first by attempting to remotely manoeuvre them (for a maximum of \textsf{nAttempts} attempts), and then by performing a hard reset. The first approach has a low cost but only a medium probability of success, while the hard reset has a high cost and higher probability of success. We model the supervisor's operation using a DTMC $m_\mathit{sup
}$ augmented with reward structures enabling the verification of the expected number of failed robots and supervisor-intervention cost across all missions performed by the robot fleet. Figure~\ref{fig:results}c shows how these RoboFleet properties vary when the external parameter \textsf{nAttempts} is varied between 1 and 4 for a fleet with $N=2$ robots.

\smallskip\noindent
\textbf{Dynamic power management (DPM) for foreign exchange (FX) system.} The DPM-FX case study considers the FX service-based system from~\cite{DBLP:conf/kbse/GerasimouTC15,DBLP:journals/tse/FangCGA23}, which performs automated financial transactions in the foreign exchange market, and the database used by the FX services tasked with the sophisticated market analyses behind this system's decision making. To reduce power usage, the  hard disk storing this large database uses a DPM component with the characteristics from~\cite{KwiatkowskaNP07}. The role of this component is to switch the hard disk to a low-power ``sleep'' mode of operation when the disk is idle (with a configurable probability \texttt{pIdle2Sleep}). We model the joint behaviour of FX and DPM using two co-dependent stochastic models: a DTMC $m_\mathit{fx}$ (taken from~\cite{DBLP:journals/tse/FangCGA23}) and a CTMC $m_\mathit{dpm}$ (adapted from~\cite{KwiatkowskaNP07}), respectively. In this co-dependency, the average length of DPM's queue of disk operations \textsf{avrQueueDiskOps} influences the number of transactions that can be performed by FX, and therefore the number of disk operations \textsf{diskOps} that FX sends to the database managed by DPM. We use the multi-model stochastic system formed by these two models and their interdependencies (Figure~\ref{fig:case-studies}) to verify how the expected execution time of the FX workflow varies for different predefined values of the external DPM parameter \textsf{pIdle2Sleep} (Figure~\ref{fig:results}d).  

\smallskip\noindent
\textbf{Robot coordination (RoCo).} In this case study we verify a two-robot system adapted from~\cite{KNPS19}. The two robots need to coordinate in order to avoid collision as they move concurrently across a square grid world, starting in diagonally opposite corners and tasked with reaching the opposite corner from the one where they started. In each time unit, the robots can move from their current location to a neighbouring one, including diagonally. However, these moves are subject to uncertainty: at each robot move, there is a small probability $q$ that the robot will move in a direction immediately adjacent to the planned one, because of obstacles detected by its cameras. For instance, if a robot plans to move south, then it will actually move south with probability $1-q$, south-east with probability $q/2$, and south-west with probability $q/2$. We model the behaviour of the two robots using a concurrent stochastic game $m_\mathit{rob}$ with dependency parameter $q$ obtained through the analysis of a DTMC model $m_\mathit{od}$ of the camera-based obstacle detection component of the robots. The DTMC models each robot's obstacle-detection workflow before each move. This workflow includes several stages in which increasingly accurate but more energy consuming and computationally intensive methods are used to conservatively determine whether the location that the robot plans to move to next contains an obstacle. We use the multi-model
stochastic system formed by these two models to verify (Figure~\ref{fig:results}e) how the combined probability $\textsf{pSucc1}+\textsf{pSucc2}$ that the robots reach their destinations without collision within $k$ moves on a $7\times 7$ grid depend on the actual probability $\textsf{pObstacle}$ that a grid cell is occupied by an obstacle.

%includes an FX system for the \textcolor{red}{currency exchange} executed in either a technical or a fundamental analysis mode, each with different trade-offs between disk operations and analysis quality. The required operations are sent to a server for processing. To avoid the server overflow, a fraction of the number of disk operations in technical analysis (analogous to fundamental analysis) is sent at a time, given by: $disk\_ops\_Technical/avr\_num\_disk\_ops\_remain\_in\_queue$ where the denominator represents the average operations waiting in the server's queue. 
%To reduce power consumption, the FX system runs on the same server as the DPM system~\cite{KwiatkowskaNP07}. The Service Requester Queue (SRQ) in the DPM receives requests at a rate of $request=1/(disk\_ops * 2)$ assuming two FX system executions per time unit and that $disk\_ops$ represents the expected total number of operations from the FX system.

\smallskip\noindent
\textbf{Discussion.} The case studies presented in this section show that our ULTIMATE approach and verification tool are applicable to multi-model stochastic systems with a diversity of heterogeneous model types, verified properties and dependency patterns, none of which is supported by existing PMC techniques and tools. In three of the case studies (RAD---see Figure~\ref{fig:RAD-models}, and SMD and DPM-FX---see Figure~\ref{fig:case-studies}), these dependency patterns included co-dependencies (i.e., circular dependencies) that are very challenging to resolve. 

ULTIMATE handled these co-dependencies successfully using one of the two methods from the description of Algorithm~\ref{algo:verification}. For the case studies with co-dependencies within strongly connected components comprising only DTMCs (i.e., RAD and SMD), the co-dependencies were handled automatically by using parametric model checking to assemble a system of polynomial equations that were then solved numerically to obtain the values of the co-dependent dependency parameters. For the DPM-FX case study, whose circular dependency is between a DTMC and a CTMC, the co-dependent parameters were estimated (also automatically) using Powell's derivative-free numerical optimisation method~\cite{powell2007view}. Our ULTIMATE tool has the in-built capability to select the appropriate method for handling each type of co-dependency.

The ULTIMATE tool also provides an intuitive, use friendly graphical user interface for assembling a multi-model stochastic system from its component models, and defining their dependency and external parameters. A demonstration video illustrating this functionality is available in our online repository~\cite{ULTIMATE-repo-non-anonymised}. 

The ULTIMATE verification times depend on those of the model checkers and solvers invoked by Algorithm~\ref{algo:verification}. Given the efficiency of modern probabilistic model checkers, all properties from Table~\ref{table:case-studies} could be verified within seconds on a standard laptop. 

% tool: intuitive GUI user-friendly tool, define the parameters easy, verification as well as see the detailed results in a log that can be saved for future inspection (video in repo)

\smallskip
\noindent
\textbf{Threats to validity.}
We limit \textbf{construct validity threats} that could arise due to simplifications and assumptions from the adopted experimental methodology by using four case studies of complex systems from a diverse set of application domains, and, where appropriate, obtaining models and their properties from the research literature~\cite{KwiatkowskaNP07,DBLP:conf/kbse/GerasimouTC15,KNPS19}.

We reduce \textbf{internal validity threats} that could introduce bias when determining cause-effect relationships in the experimental study by cross-validating the developed models independently using multiple probabilistic model checkers. 
Moreover, the multi-model stochastic system specifications were manually reviewed and corroborated by at least two researchers independently. 
Finally, we enable replication and verification of our findings by making all case studies and experimental results publicly available online.

We mitigate \textbf{external validity threats} that could affect the generalisability of our findings by demonstrating the applicability of ULTIMATE using a diverse combination of probabilistic models (see Figures~\ref{fig:RAD-models} and~\ref{fig:case-studies}) that represent systems from different application domains.
Also, ULTIMATE uses the high-level modelling language provided by the probabilistic model checker PRISM~\cite{KNP11} for model representation, and the probabilistic model checkers  Storm~\cite{hensel2022probabilistic} and PRISM~\cite{KNP11} within its verification engine, thus enhancing further its applicability due to the familiarity of the research community with these tools.
However, additional experiments are needed to confirm that ULTIMATE can analyse multi-model stochastic systems for systems and processes other than those employed in our evaluation.

\section{Related work \label{sec:related}}

The probabilistic model checking of interdependent stochastic models is an underexplored area of research. The only other approach proposed so far for the PMC of combinations of interdependent stochastic models has been the compositional assume-guarantee verification of probabilistic models~\cite{KNPQ10,calinescu2012compositional,feng2014learning,liu2020compositional}, which aims to improve PMC scalability. However, unlike our ULTIMATE framework, this compositional verification paradigm supports only one type of stochastic models (probabilistic automata), can only deal with very simple interdependencies among the models it verifies compositionally, and does not provide external parameter estimation capabilities. As such, this compositional assume-guarantee verification approach cannot be used to verify any of the multi-model stochastic systems from our case studies in Section~\ref{sec:evaluation}. 

Our ULTIMATE verification algorithm (Section~\ref{subsect:algorithm}) operates with the strongly connected components (SCCs) of the dependency graph induced by the multi-model stochastic system under verification. The use of SCC decomposition is well established in probabilistic and parametric model checking~\cite{abraham2010dtmc,hartmanns2023fast,hartmanns2023practitioner}, but at a different stage of the PMC verification process, and for a completely different purpose to ours. Thus, leading probabilistic model checkers including PRISM~\cite{KNP11} and Storm~\cite{hensel2022probabilistic} decompose the single stochastic models they analyse (e.g., DTMCs and MDPs) into SCCs in order to speed up their verification. In contrast, our ULTIMATE verification framework applies SCC decomposition to the dependency graph induced by a multi-model stochastic system, for the purpose of ordering the verification of its component models.

\hspace*{-1.5mm}To the best of our knowledge, ULTIMATE is the first tool-supported approach capable of jointly verifying sets of heterogeneous stochastic models with the complex types of model interdependencies illustrated by the case studies from Section~\ref{sec:evaluation}.

\section{Conclusion \label{sec:conclusion}}

We introduced ULTIMATE, a novel framework for 
% to address the limitations of existing probabilistic model checking techniques in 
the verification of heterogeneous multi-model stochastic systems with complex interdependencies. 
ULTIMATE provides a unified approach to stochastic modelling, verification, and synthesis, accommodating probabilistic and non-deterministic uncertainty, discrete and continuous time, and partial observability.
Furthermore, ULTIMATE is underpinned by a novel verification method for managing model interdependencies, automating dependency analysis, synthesis of analysis and parameter computation tasks, and the invocation of diverse verification engines, including probabilistic and parametric model checkers, numeric solvers, optimisers, and inference functions based on frequentist and Bayesian principles.
Through a comprehensive suite of case studies, we demonstrated ULTIMATE's capabilities to support the rigorous verification of complex software-intensive systems.

Our future work will focus on (1) extending the applicability of ULTIMATE to an even wider range of domains; (2) further refining its capacity to handle increasingly intricate model interactions and incorporate diverse forms of domain knowledge and data; and (3) supporting automatic parameter synthesis~\cite{DBLP:conf/kbse/GerasimouTC15,dehnert2015prophesy}.

\bibliographystyle{ACM-Reference-Format}
\bibliography{ULTIMATE}
\input{appendix}

\end{document}

%% file: appendix.tex
%\newpage
\section*{Appendix A. Stochastic models and the specification of their properties}
\label{sec:prism}

\subsection*{A.1 Discrete-time Markov chains}

% A few introductory sentences telling what DTMCs are about.

% \begin{definition}
% A discrete-time Markov chain is a tuple 
% \begin{equation}
%     \label{eq:dtmc}
%     D = (S, s_0, \mathbf{P}, AP, L),
% \end{equation}
% where \begin{itemize}
%     \item $S$ is a set of states;
%     \item $s_0 \in S$ is an initial state;
%     \item $P : S \times S \to [0, 1]$ is a transition probability matrix such that
%     \[
%     \sum_{s' \in S} P(s, s') = 1 \quad \text{for all } s \in S;
%     \]
%     \item $\mathcal{AP}$ is a set of atomic propositions;
%     \item $L : S \to 2^{\mathcal{AP}}$ is a labeling function that assigns, to each state $s \in S$, a set $L(s)$ of atomic propositions.
% \end{itemize}

% \end{definition}

\begin{definition}[Discrete time Markov chain]
A \emph{discrete-time Markov chain} (DTMC) over a set of atomic propositions $AP$ is a tuple 
\begin{equation}
  \mathcal{M}=(S,s_I,\mathbf{P},L),
  \label{eq:DTMC} 
\end{equation}
where: 
\begin{itemize}
\item $S=\{s_1,s_2,\ldots,s_n\}$ is a finite set of states and $s_I$ is the initial state;
\item $\mathbf{P}: S\times S\to [0,1]$ is a transition probability matrix such that for any states $s_i, s_j\in S$, $\mathbf{P}(s_i,s_j)$ represents the probability of transitioning to state $s_j$ from state $s_i$. For any state $s_i\in S$, $\sum_{s_j\in S} \mathbf{P} (s_i,s_j) = 1$. For simplicity, we will often use the notation $p_{ij}=\mathbf{P}(s_i,s_j)$.
\item $L : S \to 2^{AP}$ is a labelling function that assigns a set of atomic propositions
from $AP$ to each state in $S$.  
\end{itemize}
\end{definition}

\noindent
A \emph{path} $\pi$ over $\mathcal{M}$ is a possibly infinite sequence of states from $S$ such that for any adjacent states $s$ and $s'$ in $\pi$, $\mathbf{P}(s,s')>0$. The $m$-th state on a path $\pi$, $m\geq 1$, is denoted $\pi(m)$. Finally, for any state $s\in S$, $Paths^\mathcal{M}(s)$ represents the set of all infinite paths over $\mathcal{M}$ that start with state $s$. 

To compute the probability that a Markov chain~(\ref{eq:DTMC}) behaves in a specified way when in state $s\in S$, we use a \emph{probability measure} $\mathrm{Pr}_s$ defined over $Paths^\mathcal{M}(s)$ such that \cite{KatoenBook,kemeny-etal1976}:
\begin{equation}
  \label{eq:prob-measure}
  \begin{array}{l}
  \mathrm{Pr}_s(\{\pi\in Paths^\mathcal{M}(s)\mid \pi=s_{i_1}s_{i_2}s_{i_3}\ldots s_{i_m}\ldots\}) =\\
   \qquad\qquad\qquad\qquad\qquad\qquad\qquad p_{i_1i_2} p_{i_2i_3} \ldots p_{i_{m-1}i_m},
  \end{array}
\end{equation}
where $\{\pi\in Paths^\mathcal{M}(s)\mid \pi=s_{i_1}s_{i_2}s_{i_3}\ldots s_{i_m}\ldots\}$ represents the set of all infinite paths that start with the prefix $s_{i_1}s_{i_2}s_{i_3}\ldots s_{i_m}$ (i.e., the \emph{cylinder set} of this prefix). Further details about this probability measure and its properties are available from \cite{KatoenBook,kemeny-etal1976}.

% Note: Other concepts about DTMCs that are required to describe the semantics of their PCTL-encoded properties, e.g., paths.

To enable the analysis of a broader range of properties, DMTCs are often augmented with cost/reward structures that associate non-negative values with their states and/or transitions\cite{KwiatkowskaNP07}.

% \begin{definition}[Reward structure]
% A reward structure over a DTMC is a pair $(\mathcal{M}, \rho)$ with DTMC $\mathcal{M} = (S,s_1 P, L)$ and $\rho : S \to \mathbb{R}_{>0}$ a \textit{reward assignment function} that associates a real reward (or cost) to any state in $S$. Real number (constant) $\rho(s)$ denotes the reward earned on leaving state $s$.
% \end{definition}

\begin{definition}[Reward structure]
A reward structure for a DTMC $\mathcal{M} = (S,s_1 P, L)$ is defined by the tuple $R=$($\rho,\iota$) comprising a state reward function $\rho:S\rightarrow \mathbb{R}_{\geq0}$, and a transition reward function $\iota:S\times S\rightarrow\mathbb{R}_{\geq0}$.
\label{def:rewardStructureDTMC}
\end{definition}

\noindent
The state reward $\rho(s)$ is the reward acquired in state $s$ if the DTMC is in state $s$ for one time-step, and the transition reward $\iota(s,s')$ when a transition from $s$ to $s'$ occurs.

\medskip
\noindent
\textbf{DTMC properties.} Probabilistic model checking supports the analysis of DTMC properties specified formally in probabilistic computation tree logic (PCTL) extended with rewards.

\begin{definition}[PCTL syntax]
Let $AP$ be a set of atomic propositions and $a \in AP$, $p\in[0,1]$, $k \in \mathbb{N}$, $r\in\mathbb{R}$ and $\bowtie$ $\in \{ \geq, >, <, \leq\}$. 
Then a \emph{state formula} $\Phi$ and a \emph{path formula} $\Psi$ in probabilistic computation tree logic (PCTL) are defined by the grammar:
\begin{align}
& \Phi::=  true \;\vert\; a \;\vert\; \Phi \wedge \Phi \;\vert\; \neg \Phi \;\vert\; \mathcal{P}_{\bowtie p} [\Psi]
\label{eq:pctl} \\
& \Psi::= X \Phi \;\vert\; \Phi \mathrm{U} \Phi \;\vert\; \Phi \mathrm{U}^{\leq k} \Phi
\label{eq:pctl2}
\end{align}
and a \emph{cost/reward state formula} is defined by the grammar:
\begin{equation}
  \Phi::=  \mathcal{R}_{\bowtie r} [\mathrm{I}^{=k}] \;\vert\;
               \mathcal{R}_{\bowtie r} [\mathrm{C}^{\leq k}] \;\vert\;
               \mathcal{R}_{\bowtie r} [\mathrm{F}\Phi] \;\vert\;
               \mathcal{R}_{\bowtie r} [\mathrm{S}].
\label{eq:pctl3}
\end{equation}
\end{definition}

The semantics of PCTL is defined with a satisfaction relation $\models$ over the states $S$ and the paths $Paths^\mathcal{M}(s)$, $s\in S$, of a Markov chain~(\ref{eq:DTMC}). Thus, $s\models \Phi$ means ``$\Phi$ is satisfied in state $s$'' or ``$\Phi$ is true in state $s$''. For any state $s\in S$, we have: $s\models true$; $s \models a$ iff $a\in L(s)$; $s \models \neg \Phi$ iff $\neg (s\models \Phi)$; and $s\models \Phi_1 \wedge \Phi_2$ iff $s\models \Phi_1$ and $s\models \Phi_2$. 
A state formula $\mathcal{P}_{\bowtie p} [\Psi]$ is satisfied in a state $s$ if the probability of the future evolution of the system satisfying $\Psi$ satisfies $\bowtie p$:
\[
    \!\!\!\!s\models \mathcal{P}_{\bowtie p} (\Psi) \;\textrm{ iff }\; \mathrm{Pr}_s(\{\pi\in Paths^\mathcal{M}(s) \mid \pi \models \Psi\}) \bowtie p.
\]
The path formulae used with the probabilistic operator $\mathcal{P}$ and their semantics are:
\begin{itemize}
\item the ``next'' formula $X \Phi$, which is satisfied by a path $\pi$ iff $\Phi$ is satisfied in the next state of $\pi$ (i.e., in state $\pi(2)$);
\item the time-bounded ``until'' formula $\Phi_1 \mathrm{U}^{\leq k} \Phi_2$, which is satisfied by a path $\pi$ iff $\Phi_1$ is satisfied in each of the first $x$ states of $\pi$ for some $x< k$, and $\Phi_2$ is satisfied in the $(x+1)$-th state of $\pi$;
\item the unbounded ``until'' formula $\Phi_1 \mathrm{U} \Phi_2$, which is satisfied by a path $\pi$ iff $\Phi_1$ is true in each of the first $x\!>\!0$ states of $\pi$, and $\Phi_2$ is true in the $(x\!+\!1)$-th state of $\pi$.
\end{itemize}

\noindent
The notation $\mathrm{F}^{\leq k}\Phi\equiv \mathit{true} \mathrm{U}^{\leq k} \Phi$ and $\mathrm{F}\Phi\equiv \mathit{true} \mathrm{U} \Phi$ is used when the first part of a bounded until and until formula, respectively, is $\mathit{true}$. 

A state $s$ satisfies $\mathcal{R}_{\bowtie r}[\mathrm{C}^{\leq k}]$ if, from $s$, the expected reward cumulated after $k$ time-steps satisfies $\bowtie r$; $\mathcal{R}_{\bowtie r}[\mathrm{I}^{= k}]$ is true if, from $s$, the expected state reward at time step $k$ is $\bowtie r$; and $\mathcal{R}_{\bowtie r}[\mathrm{F} \Phi]$ is true if from state $s$ the expected reward cumulated before a state satisfying $\Phi$ is reached meets the bound $\bowtie r$~\cite{KwiatkowskaNP07}.

\subsection*{A.2 Continuous-time Markov chains}
Continuous-time Markov chains are mathematical models for continuous-time stochastic processes over countable state spaces.
\begin{definition}[Continues time Markov chain]
    A Continuous-Time Markov Chain (CTMC) is formally defined as a tuple 
$C = (S, s_I, R, L)$, 
where:
\begin{itemize}
    \item $S$ is a finite set of states (state space),
    \item $s_I \in S$ is the initial state,
    \item $R : S \times S \to \mathbb{R}_{\geq 0}$ is the transition rate matrix specifying transition rates between states, and
    \item $L : S \to 2^{AP}$ is a labeling function that maps each state to a set of atomic propositions.
\end{itemize}
\end{definition}

The transition rate matrix $R : S \times S \to \mathbb{R}_{\geq 0}$ assigns rates $R(s, s')$ for transitions between states $s$ and $s'$, where $R(s, s') > 0$ indicates a possible transition. The time until a transition is exponentially distributed with parameter $R(s, s')$, and the probability of a transition occurring within $t$ time units is $1 - e^{-R(s, s')t}$. When multiple transitions are possible, the first to occur determines the next state, a phenomenon known as race condition. The time spent in a state before transitioning is the minimum of multiple exponential distributions, with the total exit rate $E(s) = \sum_{s' \in S \setminus \{s\}} R(s, s')$. A state $s$ is absorbing if $E(s) = 0$, meaning it has no outgoing transitions. The probability of leaving a state $s$ within time $t$ is $1 - e^{-E(s)t}$.

\medskip
\noindent
\textbf{CTMC properties.} 
CTMCs support the analysis of properties expressed in \emph{continuous stochastic logic} (CSL) \cite{aziz1996}, which is a temporal logic with the syntax defined below.
\begin{definition}[CSL syntax]
Let AP be a set of atomic propositions, $a\!\in\!AP \mbox{, } p\! \in\! [0,1] \mbox{, } I$  an interval in  $\mathbb{R} \mbox{ and } \bowtie\!\! \mbox{}  \in \{\geq, >, <, \leq \}$. Then a state formula $\Phi$ and a path formula $\Psi$ in continuous stochastic logic are defined by the following grammar:
\begin{equation}
\begin{array}{l}
\Phi ::= true\, |\,  a\, |\, \Phi \wedge \Phi\, |\, \neg\Phi\, |\, P_{\bowtie p}[\Psi]\, |\, \mathcal{S}_{\bowtie p}[\Phi]\\
\Psi ::= X \Phi\, |\, \Phi U^{I} \Phi  
\end{array}.
\end{equation}

\noindent
CSL formulae are interpreted over a CTMC whose states are \emph{labelled} with atomic propositions from $AP$ by a function $L$. The (\emph{transient-state}) \emph{probabilistic operator} $P$ and the \emph{steady-state operator} $\mathcal{S}$ define bounds on the probability of system evolution. \emph{Next path formulae} $X \Phi$ and \emph{until path formulae} $\Phi_1 U^{I} \Phi_2$ can occur only inside the probabilistic operator $P$. 

The semantics of CSL is defined with a satisfaction relation $\models$ over the states $s\!\in\! S$ and the paths $\omega\!\in\! \mathit{Paths}^\mathcal{M}$ of a CTMC \cite{Baier2000HHK,KwiatkowskaNP07}. The semantics is defined recursively by:
\[
\!
\begin{array}{ll}
s\models \mathit{true} & \!\forall s\in S\\
s\models a & \!\textrm{iff } \!a\in L(s)\\
s\models \Phi_1\wedge \Phi_2 & \!\textrm{iff } s\models \Phi_1 \wedge s\models \Phi_2 \\
s\models \neg \Phi & \!\textrm{iff } \neg(s\models \Phi) \\
s\models P_{\bowtie p}[\Psi] & \!\textrm{iff } \mathit{Pr}_s \{\omega\in\mathit{Paths}^\mathcal{M}\mid \omega\models \Psi\}\bowtie p \\
\omega \models X \Phi & \!\textrm{iff } \omega=s_1t_1s_2\ldots \wedge s_2\models \Phi\\
\omega \models \Phi_1 U^{I} \Phi_2 & \!\textrm{iff } \exists t\!\in\! I. (\forall t'\!\in\![0,t).\, \omega@t'\!\models\! \Phi_1)\wedge \omega@t\!\models\! \Phi_2
\end{array}
\]
where a formal definition for the probability measure $\mathit{Pr}_s$ on paths starting in state $s$ is available in \cite{Baier2000HHK,KwiatkowskaNP07}.
\end{definition}
\subsection*{A.3 Markov decision processes}

% \begin{definition}
%     A Markov decision process (MDP) is a tuple $\mathcal{M}=(S,s_I,Act,\Delta,L)$, where $S$ is a finite set of states, $s_1$ the initial state and $L$ the labelling function as in a DTMC; $Act$ is a nonempty set of actions; and $\Delta:S\times A\rightarrow Dist(S)$ is a partial probabilistic transition function from state-action pairs to discrete probability distributions over S.
% \end{definition}
Markov decision processes generalise DTMCs with the ability to model nondeterminism~\cite{forejt2011automated}.

\begin{definition}[Markov decision process]\label{def:MDP}
A Markov decision process (MDP) over a set of atomic propositions \(\mathit{AP}\) is a tuple \(\mathcal{M} = (S, s_I, A, \Delta, L, R)\), where:

\begin{itemize}
    \item \(S\): A finite set of states.
    \item \(s_I \in S\): The initial state.
    \item \(A \neq \emptyset\): A finite set of actions.
    \item \(\Delta : S \times A \rightarrow \mathit{Dist}(S)\): A partial probabilistic transition function that maps state-action pairs to discrete probability distributions over \(S\).
    \item \(L: S \rightarrow 2^{\mathit{AP}}\): A labelling function that maps each state to a set of atomic propositions.
    \item \(R\): A reward function, defined as a tuple $R=(r_{state},r_{action})$ comprising a state reward function $r_{state}:S\rightarrow \mathbb{R}_{\geq 0}$ and an action reward function $r_{action}:S\times A\rightarrow \mathbb{R}_{\geq 0}$.
\end{itemize} %Finally, $\mathcal{D}(X)$ denotes the set of discrete probability distributions over finite set $X$.
\end{definition}

In each state $s \in S$, the set of actions $a\in A$ for which $\Delta(s,a)$ is defined contains the actions \emph{enabled} in state $s$, and is denoted by $A(s)$. 
The choice of which action from $A(s)$ to take in every state $s$ is assumed to be nondeterministic. 
We reason about the behaviour of MDPs using policies. 
A policy resolves the nondeterministic choices of an MDP, selecting the action taken in every state. 
MDP policies can be classified into infinite-memory, finite-memory and memoryless policies (depending on whether the action selected in a state depends on all, a finite number, or none of the previously visited states and on the actions selected in those states). Memoryless policies can be further classified into deterministic (when the same action is selected each time when a state is reached) and randomised (when the action selected in a state is given by a discrete probability distribution over the feasible actions). 

\begin{definition}[MDP policy]
\label{def:adversary}
A (deterministic memoryless) policy of an MDP is a function $\sigma: S \rightarrow A$ that maps each state $s\in S$ to an action from $A(s)$.
\end{definition}

\medskip
\noindent
\textbf{MDP properties.} 
MDPs support the analysis of properties expressed in PCTL extended with rewards based on the syntax defined in~\eqref{eq:pctl}--\eqref{eq:pctl3}. 
Replacing  $\bowtie\!p$ from~\eqref{eq:pctl} (or $\bowtie\!r$ from~\eqref{eq:pctl3}) with $min\!=?$ or $max\!=?$ 
enables calculating the minimum/maximum probability (or reward) over all MDP policies. %, 
% enables quantifying the minimum/maximum probability over all policies and path formula $\Psi$; similar reasoning applies for rewards.
For a full description of the PCTL semantics for MDPs, see~\cite{hansson-jonsson1994}.

% Key properties of MDPs are the probability of reaching a target and the expected reward cumulated until this occurs (where we assume that the expected value is infinite if there is a non-zero probability of the target not being reached). 

% Let \(\Sigma_M\) denote the set of all strategies of \(M\). Let \(S\) denote the target set of states. Under a specific strategy \(\sigma\) of MDP \(M\), we denote this property by $Pr_{M}^{\sigma}(F S)$. The optimal (minimum or maximum) values $Pr_{M}^{\text{opt}}(F S)$, where \(\text{opt} \in \{\text{min}, \text{max}\}\) are computed for an MDP \(M\) as follows:

% \begin{equation}
%     Pr_{\text{min}}^M(F S) \coloneqq \inf_{\sigma \in \Sigma_M} Pr_\sigma^M(F S),
% \end{equation}

% \begin{equation}
%     Pr_{\text{max}}^M(F S) \coloneqq \sup_{\sigma \in \Sigma_M} Pr_\sigma^M(F S).
% \end{equation}

\subsection*{A.4 Probabilistic automata}

PAs are a generalisation of MDPs where a transition from a (state, action) pair can map to multiple probability distributions over the set of states $S$~\cite{segala1995probabilistic}. 

\begin{definition}[Probabilistic automaton]
    A \emph{probabilistic automaton} (PA) is defined by the tuple $\mathcal{M}=(S,s_I,A,\delta,L)$, where:
    \begin{itemize}
        \item ($S,s_I,A$) is a set of states, an initial state and a set of actions as in an MDP;
        \item $\delta\subseteq S\times A \times Dist(S)$ is a probabilistic transition relation where $Dist(S)$ denotes the set of discrete distribution functions over $S$.
    \end{itemize}
    % . As in an MDP, $S$ is a finite set of states, $s_I\in S$ the initial state, $A$ the action alphabet, and $L$ the labelling function; probabilistic transition relations are defined as $\delta\subseteq S\times A \times Dist(S)$, where $Dist(S)$ denotes the set of discrete distribution functions over $S$.
    \label{def:PA}
\end{definition}

A PA transition is defined as 
\(\delta \subseteq S \times \mathit{Dist}(\mathit{A} \times S \cup \{\tau\})\) in~\cite{segala1995probabilistic}, 
where a probabilistic choice of actions is allowed. Here, \(\tau\) represents a deadlock, meaning no further transitions are possible. Since model checkers like PRISM resolve deadlocks by adding self-loops to such states~\cite{forejt2011automated}, we do not consider \(\tau\) in our analysis. 

Furthermore,~\cite{segala1995probabilistic} focuses primarily on \textit{simple PAs} for most of its analysis. 
In a simple PA, a transition \((s, \mathit{Dist}(\mathit{A} \times S))\) has probability distributions associated with a single action. 
This is represented as \((s, a, \mathit{Dist}(S))\), as described in Definition~\ref{def:PA}. Finally, reward structures can be added to PAs as for MDPs.

% \begin{definition}
%     A \emph{probabilistic automata} (PA) is defined by the tuple $\mathcal{M}=(S,s_1,A,\delta,L)$. As in an MDP, $S$ is a finite set of states, $s_1\in S$ the initial state, $A$ the action alphabet, and $L$ the labelling function. Probabilistic transition relations are defined as $\delta\subseteq S\times Dist(A \times S \cup \{\tau\})$, where $Dist(S)$ denotes the set of discrete distribution functions over $S$, and $\tau$ is defined as a deadlock.
% \end{definition}

% \textcolor{red}{On Monterey's definition}, this differs from Segala's one as $\tau$ is defined as an action not an action-state pair, and the prob. distribution is only applied to states:

% \vspace{2mm}
% \noindent
%     \textcolor{red}{
%     \textbf{Definition in Monterey:}
%     A \emph{probabilistic automata} (PA) is defined by the tuple $\mathcal{M}=(S,s_1,A,\delta,L)$. As in an MDP, $S$ is a finite set of states, $s_1\in S$ the initial state, $A$ the action alphabet, and $L$ the labelling function. Probabilistic transition relations are defined as $\delta\subseteq S\times (A \cup \{\tau\}) \times Dist(S)$, where $Dist(S)$ denotes the set of discrete distribution functions over $S$.
%     }

\subsection*{A.5 Partially observable MDPs}
\begin{definition}[Partially observable MDP]
    A POMDP is a tuple $M = (S, s_I,$ $A, P, L, R, O, \text{obs})$ where:
\begin{itemize}
    \item $(S, s_I, A, P, L, R)$ is an MDP;
    \item $O$ is a finite set of observations;
    \item $\text{obs} : S \to O$ is a labelling of states with observations;
\end{itemize}
such that, for any states $s, s' \in S$ with $\text{obs}(s) = \text{obs}(s')$, their available actions must be identical, i.e., $A(s) = A(s')$. 
\end{definition}

The current state $s$ of a POMDP cannot be directly determined, only the corresponding observation $\text{obs}(s) \in O$ is known~\cite{NPZ17}. 

\begin{definition}[Observation-based strategy]
   A strategy of a POMDP \(M = (S, s_I, A, P, L, R, O, \text{obs})\) is a function \(\sigma : \text{FPaths}_M \to \text{Dist}(A)\) such that:
\begin{itemize}
    \item \(\sigma\) is a strategy of the MDP \((S, s_I, A, P, L, R)\);
    \item For any paths \(\pi = s_0 \xrightarrow{a_0} s_1 \xrightarrow{a_1} \cdots s_n\) and \(\pi' = s_0 \xrightarrow{a_0'} s_1 \xrightarrow{a_1'} \cdots s_n\) satisfying \(\text{obs}(s_i) = \text{obs}(s_i')\) and \(a_i = a_i'\) for all \(i\), we have \(\sigma(\pi) = \sigma(\pi')\).
\end{itemize}
\end{definition}

\medskip
\noindent
\textbf{POMDP properties.} 
POMDPs, similar to MDPs, use PCTL extended with rewards based on the syntax defined in~\eqref{eq:pctl}--\eqref{eq:pctl3} for the analysis of properties. 
Replacing  $\bowtie\!p$ from~\eqref{eq:pctl} (or $\bowtie\!r$ from~\eqref{eq:pctl3}) with $min\!=?$ or $max\!=?$ 
enables quantifying the minimum/maximum probability (or reward) over all POMDP policies. %, 
Further details about the PCTL semantics over POMDPs are available in~\cite{NPZ17}.

% Let \(\Sigma_M\) denote the set of all (observation-based) strategies of \(M\). Let \(O\) denote the target (i.e., a set of observations). 

%  Under a specific strategy \(\sigma\) of POMDP \(M\), %we denote this property by $\mathbb{E}_{M}^{\sigma}(F O)$.
%  The optimal (minimum or maximum) values $\mathbb{E}_{M}^{\text{opt}}(F O)$, where \(\text{opt} \in \{\text{min}, \text{max}\}\) are computed as follows:

% \begin{equation}
% \mathbb{E}_{\text{min}}^M(F O) \coloneqq \inf_{\sigma \in \Sigma_M} \mathbb{E}_\sigma^M(F O),
% \end{equation}
% \begin{equation}
%     \mathbb{E}_{\text{max}}^M(F O) \coloneqq \sup_{\sigma \in \Sigma_M} \mathbb{E}_\sigma^M(F O).
% \end{equation}

\subsection*{A.6 Stochastic games}
\begin{definition}[Stochastic game]
    A stochastic multi-player game (SG) is a tuple~\cite{Kwiatkowska18PW} 
\[
G = (\Pi, S, (S_i)_{i \in \Pi}, \overline{s}, A, \delta, L),
\]
where:
\begin{itemize}
    \item $\Pi$ is a finite set of players,
    \item $S$ is a finite set of states,
    \item $(S_i)_{i \in \Pi}$ is a partition of $S$,
    \item $\overline{s} \in S$ is an initial state,
    \item $A$ is a finite set of actions,
    \item $\delta : S \times A \to \text{Dist}(S)$ is a (partial) probabilistic transition function,
    \item $L : S \to 2^{AP}$ is a labelling function mapping states to sets of atomic propositions from a set $AP$.
\end{itemize}
\end{definition}
The game evolves from the initial state $\overline{s}$ as follows. In any state $s$, the controlling player $i$ (where $s \in S_i$) makes a non-deterministic choice between the set of enabled actions $A(s) \subseteq A$, where $A(s) = \{a \in A \mid \delta(s,a) \text{ is defined}\}$. We assume $A(s)$ is non-empty for all states $s$, ensuring no deadlock states exist in the model. Once player $i$ selects an action $a \in A(s)$, a transition to a successor state occurs randomly according to the probability distribution $\delta(s,a)$, where the probability of transitioning to state $s'$ from the current state $s$ is $\delta(s,a)(s')$.

A path through $G$ represents a possible execution as a sequence $\pi = s_0a_0s_1a_1s_2...$ where $s_i \in S$, $a_i \in A(s_i)$, and $\delta(s_i, a_i)(s_{i+1}) > 0$ for all $i \in \mathbb{N}$. The sets of finite and infinite paths starting from state $s$ are denoted as $\text{FPath}_{G,s}$ and $\text{IPath}_{G,s}$ respectively, with $\text{FPath}_G$ and $\text{IPath}_G$ representing all such paths.
\begin{definition}[Strategy]
A strategy for player $i \in \Pi$ is formally defined as $\sigma_i: (SA)^*S_i \rightarrow \text{Dist}(A)$, which maps each finite path $\lambda.s \in \text{FPath}_G$ where $s \in S_i$ to a probability distribution $\sigma_i(\lambda.s)$ over $A(s)$, with $\Sigma_i$ denoting the set of all strategies for player $i$.
\end{definition}
\begin{definition}[Memoryless strategy]
    A strategy $\sigma_i$ is called memoryless if it only considers the current state when resolving non-determinism, i.e., if
\[
\sigma_i(\lambda \cdot s) = \sigma_i(\lambda' \cdot s) \quad \text{for all paths} \quad \lambda \cdot s, \lambda' \cdot s \in \text{FPath}_G.
\]
\end{definition}
\medskip
\noindent
\textbf{SG properties.} 
In order to formally specify the desired behaviour of an SMG, the properties are defined in a logic called rPATL~\cite{chen2013automatic}\cite{Kwiatkowska18PW}.

\begin{definition}[rPATL syntax]
The syntax of the logic rPATL is given by the grammar:

\begin{align*}
\phi &::= \text{true} \mid a \mid \neg \phi \mid \phi \land \phi \mid \ll\mathcal{C}\gg \theta \\
\theta &::= P_{\bowtie p}[\psi] \mid R^r_{\bowtie x}[F^*\phi] \\
\psi &::= X \phi \mid \phi \, U^{\leq k}\, \phi \mid \phi \, U \, \phi
\end{align*}

where \(a \in \text{AP}\), \(\mathcal{C} \subseteq \Pi\), \(\bowtie \in \{<, \leq, \geq, >\}\), \(p \in \mathbb{Q} \cap [0, 1]\), \(x \in \mathbb{Q}_{\geq 0}\), \(r\) is a reward structure, \(* \in \{0, \infty, c\}\), and \(k \in \mathbb{N}\).

A property of an SMG \(G\) expressed in rPATL is a formula from the rule \(\phi\) in the syntax above. The key operator is \(\ll \mathcal{C} \gg \theta\), where \(\mathcal{C} \subseteq \Pi\) is a coalition of players from \(G\) and \(\theta\) is a quantitative objective that this set of players will aim to satisfy.

An objective \(\theta\) is a single instance of either the \(P\) or \(R\) operator: $P_{\bowtie p}[ \cdot ]$
means that the probability of some event being satisfied should meet the bound \(\bowtie p\), and
$R^r_{\bowtie x}[ \cdot ]$
means that the expected value of a specified reward measure (using reward structure \(r\)) meets the bound \(\bowtie x\).
\end{definition}
\newpage

%% file: ARXIV-verification-paper.bbl

\begin{thebibliography}{61}

%%% the Web address.

\ifx \showCODEN    \undefined \def \showCODEN     #1{\unskip}     \fi
\ifx \showDOI      \undefined \def \showDOI       #1{#1}\fi
\ifx \showISBNx    \undefined \def \showISBNx     #1{\unskip}     \fi
\ifx \showISBNxiii \undefined \def \showISBNxiii  #1{\unskip}     \fi
\ifx \showISSN     \undefined \def \showISSN      #1{\unskip}     \fi
\ifx \showLCCN     \undefined \def \showLCCN      #1{\unskip}     \fi
\ifx \shownote     \undefined \def \shownote      #1{#1}          \fi
\ifx \showarticletitle \undefined \def \showarticletitle #1{#1}   \fi
\ifx \showURL      \undefined \def \showURL       {\relax}        \fi
% invisible to TeX
\providecommand\bibfield[2]{#2}
\providecommand\bibinfo[2]{#2}
\providecommand\natexlab[1]{#1}
\providecommand\showeprint[2][]{arXiv:#2}

\bibitem[Abrah{\'a}m et~al\mbox{.}(2010)]%
        {abraham2010dtmc}
\bibfield{author}{\bibinfo{person}{Erika Abrah{\'a}m}, \bibinfo{person}{Nils
  Jansen}, \bibinfo{person}{Ralf Wimmer}, \bibinfo{person}{Joost-Pieter
  Katoen}, {and} \bibinfo{person}{Bernd Becker}.}
  \bibinfo{year}{2010}\natexlab{}.
\newblock \showarticletitle{{DTMC} model checking by {SCC} reduction}. In
  \bibinfo{booktitle}{\emph{2010 Seventh International Conference on the
  Quantitative Evaluation of Systems}}. IEEE, \bibinfo{pages}{37--46}.
\newblock


\bibitem[Alasmari et~al\mbox{.}(2022)]%
        {DBLP:journals/jss/AlasmariCPM22}
\bibfield{author}{\bibinfo{person}{Naif Alasmari}, \bibinfo{person}{Radu
  Calinescu}, \bibinfo{person}{Colin Paterson}, {and} \bibinfo{person}{Raffaela
  Mirandola}.} \bibinfo{year}{2022}\natexlab{}.
\newblock \showarticletitle{Quantitative verification with adaptive uncertainty
  reduction}.
\newblock \bibinfo{journal}{\emph{J. Syst. Softw.}}  \bibinfo{volume}{188}
  (\bibinfo{year}{2022}), \bibinfo{pages}{111275}.
\newblock
\urldef\tempurl%
\url{https://doi.org/10.1016/J.JSS.2022.111275}
\showDOI{\tempurl}


\bibitem[Alur and Henzinger(1996)]%
        {reactive}
\bibfield{author}{\bibinfo{person}{R. Alur} {and} \bibinfo{person}{T.A.
  Henzinger}.} \bibinfo{year}{1996}\natexlab{}.
\newblock \showarticletitle{Reactive modules}. In
  \bibinfo{booktitle}{\emph{Proceedings 11th Annual IEEE Symposium on Logic in
  Computer Science}}. \bibinfo{pages}{207--218}.
\newblock
\urldef\tempurl%
\url{https://doi.org/10.1109/LICS.1996.561320}
\showDOI{\tempurl}


\bibitem[Andova et~al\mbox{.}(2004)]%
        {Andova-etal2004}
\bibfield{author}{\bibinfo{person}{Suzana Andova}, \bibinfo{person}{Holder
  Hermanns}, {and} \bibinfo{person}{Joost-Pieter Katoen}.}
  \bibinfo{year}{2004}\natexlab{}.
\newblock \showarticletitle{Discrete-Time Rewards Model-Checked}.
\newblock In \bibinfo{booktitle}{\emph{FORMATS 2003}},
  \bibfield{editor}{\bibinfo{person}{K.~G. Larsen} {and}
  \bibinfo{person}{P.~Niebert}} (Eds.). \bibinfo{series}{Lecture Notes in
  Computer Science}, Vol.~\bibinfo{volume}{2791}. \bibinfo{publisher}{Springer
  Verlag}, \bibinfo{pages}{88--104}.
\newblock


\bibitem[Aziz et~al\mbox{.}(1996)]%
        {aziz1996}
\bibfield{author}{\bibinfo{person}{Adnan Aziz}, \bibinfo{person}{Kumud Sanwal},
  \bibinfo{person}{Vigyan Singhal}, {and} \bibinfo{person}{Robert Brayton}.}
  \bibinfo{year}{1996}\natexlab{}.
\newblock \showarticletitle{Verifying continuous time {M}arkov chains}. In
  \bibinfo{booktitle}{\emph{Computer Aided Verification}}. Springer,
  \bibinfo{pages}{269--276}.
\newblock


\bibitem[Baier et~al\mbox{.}(2000)]%
        {Baier2000HHK}
\bibfield{author}{\bibinfo{person}{Christel Baier}, \bibinfo{person}{Boudewijn
  Haverkort}, \bibinfo{person}{Holger Hermanns}, {and}
  \bibinfo{person}{Joost-Pieter Katoen}.} \bibinfo{year}{2000}\natexlab{}.
\newblock \showarticletitle{Model Checking Continuous-Time Markov Chains by
  Transient Analysis}. In \bibinfo{booktitle}{\emph{Computer Aided
  Verification}}, \bibfield{editor}{\bibinfo{person}{E.~Allen Emerson} {and}
  \bibinfo{person}{Aravinda~Prasad Sistla}} (Eds.).
  \bibinfo{publisher}{Springer Berlin Heidelberg}, \bibinfo{address}{Berlin,
  Heidelberg}, \bibinfo{pages}{358--372}.
\newblock


\bibitem[Baier and Katoen(2008)]%
        {KatoenBook}
\bibfield{author}{\bibinfo{person}{Christel Baier} {and}
  \bibinfo{person}{Joost{-}Pieter Katoen}.} \bibinfo{year}{2008}\natexlab{}.
\newblock \bibinfo{booktitle}{\emph{Principles of model checking}}.
\newblock \bibinfo{publisher}{{MIT} Press}.
\newblock
\showISBNx{978-0-262-02649-9}


\bibitem[Calinescu et~al\mbox{.}(2018a)]%
        {DBLP:journals/jss/CalinescuCGKP18}
\bibfield{author}{\bibinfo{person}{Radu Calinescu}, \bibinfo{person}{Milan
  Ceska}, \bibinfo{person}{Simos Gerasimou}, \bibinfo{person}{Marta
  Kwiatkowska}, {and} \bibinfo{person}{Nicola Paoletti}.}
  \bibinfo{year}{2018}\natexlab{a}.
\newblock \showarticletitle{Efficient synthesis of robust models for stochastic
  systems}.
\newblock \bibinfo{journal}{\emph{J. Syst. Softw.}}  \bibinfo{volume}{143}
  (\bibinfo{year}{2018}), \bibinfo{pages}{140--158}.
\newblock
\urldef\tempurl%
\url{https://doi.org/10.1016/J.JSS.2018.05.013}
\showDOI{\tempurl}


\bibitem[Calinescu et~al\mbox{.}(2016)]%
        {DBLP:journals/tr/CalinescuGJPRT16}
\bibfield{author}{\bibinfo{person}{Radu Calinescu}, \bibinfo{person}{Carlo
  Ghezzi}, \bibinfo{person}{Kenneth Johnson}, \bibinfo{person}{Mauro
  Pezz{\`{e}}}, {et~al\mbox{.}}} \bibinfo{year}{2016}\natexlab{}.
\newblock \showarticletitle{Formal Verification With Confidence Intervals to
  Establish Quality of Service Properties of Software Systems}.
\newblock \bibinfo{journal}{\emph{{IEEE} Trans. Reliab.}} \bibinfo{volume}{65},
  \bibinfo{number}{1} (\bibinfo{year}{2016}), \bibinfo{pages}{107--125}.
\newblock
\urldef\tempurl%
\url{https://doi.org/10.1109/TR.2015.2452931}
\showDOI{\tempurl}


\bibitem[Calinescu et~al\mbox{.}(2011)]%
        {DBLP:journals/tse/CalinescuGKMT11}
\bibfield{author}{\bibinfo{person}{Radu Calinescu}, \bibinfo{person}{Lars
  Grunske}, \bibinfo{person}{Marta~Z. Kwiatkowska}, \bibinfo{person}{Raffaela
  Mirandola}, {and} \bibinfo{person}{Giordano Tamburrelli}.}
  \bibinfo{year}{2011}\natexlab{}.
\newblock \showarticletitle{Dynamic QoS Management and Optimization in
  Service-Based Systems}.
\newblock \bibinfo{journal}{\emph{{IEEE} Trans. Software Eng.}}
  \bibinfo{volume}{37}, \bibinfo{number}{3} (\bibinfo{year}{2011}),
  \bibinfo{pages}{387--409}.
\newblock
\urldef\tempurl%
\url{https://doi.org/10.1109/TSE.2010.92}
\showDOI{\tempurl}


\bibitem[Calinescu et~al\mbox{.}(2024)]%
        {10496502}
\bibfield{author}{\bibinfo{person}{Radu Calinescu}, \bibinfo{person}{Calum
  Imrie}, \bibinfo{person}{Ravi Mangal}, {et~al\mbox{.}}}
  \bibinfo{year}{2024}\natexlab{}.
\newblock \showarticletitle{Controller Synthesis for Autonomous Systems with
  Deep-Learning Perception Components}.
\newblock \bibinfo{journal}{\emph{IEEE Trans. Software Eng.}}
  (\bibinfo{year}{2024}), \bibinfo{pages}{1--22}.
\newblock
\urldef\tempurl%
\url{https://doi.org/10.1109/TSE.2024.3385378}
\showDOI{\tempurl}


\bibitem[Calinescu et~al\mbox{.}(2012)]%
        {calinescu2012compositional}
\bibfield{author}{\bibinfo{person}{Radu Calinescu}, \bibinfo{person}{Shinji
  Kikuchi}, {and} \bibinfo{person}{Kenneth Johnson}.}
  \bibinfo{year}{2012}\natexlab{}.
\newblock \showarticletitle{Compositional reverification of probabilistic
  safety properties for large-scale complex IT systems}. In
  \bibinfo{booktitle}{\emph{Monterey Workshop}}. Springer,
  \bibinfo{pages}{303--329}.
\newblock


\bibitem[Calinescu and Kwiatkowska(2009)]%
        {DBLP:conf/icse/CalinescuK09}
\bibfield{author}{\bibinfo{person}{Radu Calinescu} {and}
  \bibinfo{person}{Marta~Z. Kwiatkowska}.} \bibinfo{year}{2009}\natexlab{}.
\newblock \showarticletitle{Using quantitative analysis to implement autonomic
  {IT} systems}. In \bibinfo{booktitle}{\emph{31st International Conference on
  Software Engineering ({ICSE})}}. \bibinfo{publisher}{{IEEE}},
  \bibinfo{pages}{100--110}.
\newblock
\urldef\tempurl%
\url{https://doi.org/10.1109/ICSE.2009.5070512}
\showDOI{\tempurl}


\bibitem[Calinescu et~al\mbox{.}(2018b)]%
        {DBLP:journals/tse/CalinescuWGIHK18}
\bibfield{author}{\bibinfo{person}{Radu Calinescu}, \bibinfo{person}{Danny
  Weyns}, \bibinfo{person}{Simos Gerasimou}, {et~al\mbox{.}}}
  \bibinfo{year}{2018}\natexlab{b}.
\newblock \showarticletitle{Engineering Trustworthy Self-Adaptive Software with
  Dynamic Assurance Cases}.
\newblock \bibinfo{journal}{\emph{{IEEE} Trans. Software Eng.}}
  \bibinfo{volume}{44}, \bibinfo{number}{11} (\bibinfo{year}{2018}),
  \bibinfo{pages}{1039--1069}.
\newblock
\urldef\tempurl%
\url{https://doi.org/10.1109/TSE.2017.2738640}
\showDOI{\tempurl}


\bibitem[{\v{C}}e{\v{s}}ka et~al\mbox{.}(2019)]%
        {vcevska2019shepherding}
\bibfield{author}{\bibinfo{person}{Milan {\v{C}}e{\v{s}}ka},
  \bibinfo{person}{Nils Jansen}, \bibinfo{person}{Sebastian Junges}, {and}
  \bibinfo{person}{Joost-Pieter Katoen}.} \bibinfo{year}{2019}\natexlab{}.
\newblock \showarticletitle{Shepherding hordes of Markov chains}. In
  \bibinfo{booktitle}{\emph{25th International Conference on Tools and
  Algorithms for the Construction and Analysis of Systems (TACAS)}}. Springer,
  \bibinfo{pages}{172--190}.
\newblock


\bibitem[Chen et~al\mbox{.}(2013)]%
        {chen2013automatic}
\bibfield{author}{\bibinfo{person}{Taolue Chen}, \bibinfo{person}{Vojt{\v{e}}ch
  Forejt}, \bibinfo{person}{Marta Kwiatkowska}, \bibinfo{person}{David Parker},
  {and} \bibinfo{person}{Aistis Simaitis}.} \bibinfo{year}{2013}\natexlab{}.
\newblock \showarticletitle{Automatic verification of competitive stochastic
  systems}.
\newblock \bibinfo{journal}{\emph{Formal Methods in System Design}}
  \bibinfo{volume}{43} (\bibinfo{year}{2013}), \bibinfo{pages}{61--92}.
\newblock


\bibitem[Daws(2005)]%
        {Daws:2004:SPM:2102873.2102899}
\bibfield{author}{\bibinfo{person}{Conrado Daws}.}
  \bibinfo{year}{2005}\natexlab{}.
\newblock \showarticletitle{Symbolic and Parametric Model Checking of
  Discrete-time {M}arkov Chains}. In \bibinfo{booktitle}{\emph{First
  International Conference on Theoretical Aspects of Computing (ICTAC)}}.
  \bibinfo{pages}{280--294}.
\newblock


\bibitem[De~Alfaro(1998)]%
        {de1998formal}
\bibfield{author}{\bibinfo{person}{Luca De~Alfaro}.}
  \bibinfo{year}{1998}\natexlab{}.
\newblock \emph{\bibinfo{title}{Formal verification of probabilistic systems}}.
\newblock \bibinfo{thesistype}{Ph.\,D. Dissertation}. \bibinfo{school}{Stanford
  University}.
\newblock


\bibitem[Dehnert et~al\mbox{.}(2015)]%
        {dehnert2015prophesy}
\bibfield{author}{\bibinfo{person}{Christian Dehnert},
  \bibinfo{person}{Sebastian Junges}, \bibinfo{person}{Nils Jansen},
  \bibinfo{person}{Florian Corzilius}, \bibinfo{person}{Matthias Volk},
  \bibinfo{person}{Harold Bruintjes}, \bibinfo{person}{Joost-Pieter Katoen},
  {and} \bibinfo{person}{Erika {\'A}brah{\'a}m}.}
  \bibinfo{year}{2015}\natexlab{}.
\newblock \showarticletitle{Prophesy: A probabilistic parameter synthesis
  tool}. In \bibinfo{booktitle}{\emph{Computer Aided Verification: 27th
  International Conference, CAV 2015, San Francisco, CA, USA, July 18-24, 2015,
  Proceedings, Part I 27}}. Springer, \bibinfo{pages}{214--231}.
\newblock


\bibitem[Epifani et~al\mbox{.}(2009)]%
        {epifani2009model}
\bibfield{author}{\bibinfo{person}{Ilenia Epifani}, \bibinfo{person}{Carlo
  Ghezzi}, \bibinfo{person}{Raffaela Mirandola}, {and}
  \bibinfo{person}{Giordano Tamburrelli}.} \bibinfo{year}{2009}\natexlab{}.
\newblock \showarticletitle{Model evolution by run-time parameter adaptation}.
  In \bibinfo{booktitle}{\emph{31st International Conference on Software
  Engineering}}. IEEE, \bibinfo{pages}{111--121}.
\newblock


\bibitem[Fang et~al\mbox{.}(2023)]%
        {DBLP:journals/tse/FangCGA23}
\bibfield{author}{\bibinfo{person}{Xinwei Fang}, \bibinfo{person}{Radu
  Calinescu}, \bibinfo{person}{Simos Gerasimou}, {and} \bibinfo{person}{Faisal
  Alhwikem}.} \bibinfo{year}{2023}\natexlab{}.
\newblock \showarticletitle{Fast Parametric Model Checking With Applications to
  Software Performability Analysis}.
\newblock \bibinfo{journal}{\emph{{IEEE} Trans. Software Eng.}}
  \bibinfo{volume}{49}, \bibinfo{number}{10} (\bibinfo{year}{2023}),
  \bibinfo{pages}{4707--4730}.
\newblock
\urldef\tempurl%
\url{https://doi.org/10.1109/TSE.2023.3313645}
\showDOI{\tempurl}


\bibitem[Feng(2014)]%
        {feng2014learning}
\bibfield{author}{\bibinfo{person}{Lu Feng}.} \bibinfo{year}{2014}\natexlab{}.
\newblock \emph{\bibinfo{title}{On learning assumptions for compositional
  verification of probabilistic systems}}.
\newblock \bibinfo{thesistype}{Ph.\,D. Dissertation}.
  \bibinfo{school}{University of Oxford}.
\newblock


\bibitem[Feng et~al\mbox{.}(2011)]%
        {feng2011learning}
\bibfield{author}{\bibinfo{person}{Lu Feng}, \bibinfo{person}{Tingting Han},
  \bibinfo{person}{Marta Kwiatkowska}, {and} \bibinfo{person}{David Parker}.}
  \bibinfo{year}{2011}\natexlab{}.
\newblock \showarticletitle{Learning-based compositional verification for
  synchronous probabilistic systems}. In \bibinfo{booktitle}{\emph{9th
  International Symposium on Automated Technology for Verification and Analysis
  (ATVA)}}. Springer, \bibinfo{pages}{511--521}.
\newblock


\bibitem[Forejt et~al\mbox{.}(2011)]%
        {forejt2011automated}
\bibfield{author}{\bibinfo{person}{Vojt{\v{e}}ch Forejt},
  \bibinfo{person}{Marta Kwiatkowska}, \bibinfo{person}{Gethin Norman}, {and}
  \bibinfo{person}{David Parker}.} \bibinfo{year}{2011}\natexlab{}.
\newblock \showarticletitle{Automated verification techniques for probabilistic
  systems}.
\newblock \bibinfo{journal}{\emph{Formal Methods for Eternal Networked Software
  Systems: 11th International School on Formal Methods for the Design of
  Computer, Communication and Software Systems, SFM 2011, Bertinoro, Italy,
  June 13-18, 2011. Advanced Lectures 11}} (\bibinfo{year}{2011}),
  \bibinfo{pages}{53--113}.
\newblock


\bibitem[Gainer et~al\mbox{.}(2018)]%
        {gainer2018accelerated}
\bibfield{author}{\bibinfo{person}{Paul Gainer}, \bibinfo{person}{Ernst~Moritz
  Hahn}, {and} \bibinfo{person}{Sven Schewe}.} \bibinfo{year}{2018}\natexlab{}.
\newblock \showarticletitle{Accelerated model checking of parametric {M}arkov
  chains}. In \bibinfo{booktitle}{\emph{International Symposium on Automated
  Technology for Verification and Analysis (ATVA)}}. Springer,
  \bibinfo{pages}{300--316}.
\newblock


\bibitem[Gerasimou et~al\mbox{.}(2018)]%
        {DBLP:journals/ase/GerasimouCT18}
\bibfield{author}{\bibinfo{person}{Simos Gerasimou}, \bibinfo{person}{Radu
  Calinescu}, {and} \bibinfo{person}{Giordano Tamburrelli}.}
  \bibinfo{year}{2018}\natexlab{}.
\newblock \showarticletitle{Synthesis of probabilistic models for
  quality-of-service software engineering}.
\newblock \bibinfo{journal}{\emph{Autom. Softw. Eng.}} \bibinfo{volume}{25},
  \bibinfo{number}{4} (\bibinfo{year}{2018}), \bibinfo{pages}{785--831}.
\newblock
\urldef\tempurl%
\url{https://doi.org/10.1007/S10515-018-0235-8}
\showDOI{\tempurl}


\bibitem[Gerasimou et~al\mbox{.}(2015)]%
        {DBLP:conf/kbse/GerasimouTC15}
\bibfield{author}{\bibinfo{person}{Simos Gerasimou}, \bibinfo{person}{Giordano
  Tamburrelli}, {and} \bibinfo{person}{Radu Calinescu}.}
  \bibinfo{year}{2015}\natexlab{}.
\newblock \showarticletitle{Search-Based Synthesis of Probabilistic Models for
  Quality-of-Service Software Engineering {(T)}}. In
  \bibinfo{booktitle}{\emph{30th {IEEE/ACM} International Conference on
  Automated Software Engineering (ASE)}},
  \bibfield{editor}{\bibinfo{person}{Myra~B. Cohen}, \bibinfo{person}{Lars
  Grunske}, {and} \bibinfo{person}{Michael Whalen}} (Eds.).
  \bibinfo{publisher}{{IEEE} Computer Society}, \bibinfo{pages}{319--330}.
\newblock
\urldef\tempurl%
\url{https://doi.org/10.1109/ASE.2015.22}
\showDOI{\tempurl}


\bibitem[Ghezzi and Sharifloo(2013)]%
        {ghezzi2013model}
\bibfield{author}{\bibinfo{person}{Carlo Ghezzi} {and}
  \bibinfo{person}{Amir~Molzam Sharifloo}.} \bibinfo{year}{2013}\natexlab{}.
\newblock \showarticletitle{Model-based verification of quantitative
  non-functional properties for software product lines}.
\newblock \bibinfo{journal}{\emph{Information and Software Technology}}
  \bibinfo{volume}{55}, \bibinfo{number}{3} (\bibinfo{year}{2013}),
  \bibinfo{pages}{508--524}.
\newblock


\bibitem[Ghezzi and Tamburrelli(2009)]%
        {10.1007/978-3-642-02351-4_5}
\bibfield{author}{\bibinfo{person}{Carlo Ghezzi} {and}
  \bibinfo{person}{Giordano Tamburrelli}.} \bibinfo{year}{2009}\natexlab{}.
\newblock \showarticletitle{Predicting Performance Properties for Open Systems
  with KAMI}. In \bibinfo{booktitle}{\emph{Architectures for Adaptive Software
  Systems}}, \bibfield{editor}{\bibinfo{person}{Raffaela Mirandola},
  \bibinfo{person}{Ian Gorton}, {and} \bibinfo{person}{Christine Hofmeister}}
  (Eds.). \bibinfo{publisher}{Springer Berlin Heidelberg},
  \bibinfo{address}{Berlin, Heidelberg}, \bibinfo{pages}{70--85}.
\newblock


\bibitem[Hahn et~al\mbox{.}(2010)]%
        {param}
\bibfield{author}{\bibinfo{person}{Ernst~Moritz Hahn}, \bibinfo{person}{Holger
  Hermanns}, \bibinfo{person}{Bj{\"o}rn Wachter}, {and} \bibinfo{person}{Lijun
  Zhang}.} \bibinfo{year}{2010}\natexlab{}.
\newblock \showarticletitle{{PARAM: A} Model Checker for Parametric {M}arkov
  Models}. In \bibinfo{booktitle}{\emph{International Conference on Computer
  Aided Verification (CAV)}}. \bibinfo{pages}{660--664}.
\newblock


\bibitem[Hansson and Jonsson(1994)]%
        {hansson-jonsson1994}
\bibfield{author}{\bibinfo{person}{Hans Hansson} {and} \bibinfo{person}{Bengt
  Jonsson}.} \bibinfo{year}{1994}\natexlab{}.
\newblock \showarticletitle{A logic for reasoning about time and reliability}.
\newblock \bibinfo{journal}{\emph{Formal Aspects of Computing}}
  \bibinfo{volume}{6}, \bibinfo{number}{5} (\bibinfo{year}{1994}),
  \bibinfo{pages}{512--535}.
\newblock


\bibitem[Hartmanns et~al\mbox{.}(2023a)]%
        {hartmanns2023practitioner}
\bibfield{author}{\bibinfo{person}{Arnd Hartmanns}, \bibinfo{person}{Sebastian
  Junges}, \bibinfo{person}{Tim Quatmann}, {and} \bibinfo{person}{Maximilian
  Weininger}.} \bibinfo{year}{2023}\natexlab{a}.
\newblock \showarticletitle{A practitioner’s guide to {MDP} model checking
  algorithms}. In \bibinfo{booktitle}{\emph{International Conference on Tools
  and Algorithms for the Construction and Analysis of Systems}}. Springer,
  \bibinfo{pages}{469--488}.
\newblock


\bibitem[Hartmanns et~al\mbox{.}(2023b)]%
        {hartmanns2023fast}
\bibfield{author}{\bibinfo{person}{Arnd Hartmanns}, \bibinfo{person}{Bram
  Kohlen}, {and} \bibinfo{person}{Peter Lammich}.}
  \bibinfo{year}{2023}\natexlab{b}.
\newblock \showarticletitle{Fast verified {SCCs} for probabilistic model
  checking}. In \bibinfo{booktitle}{\emph{International Symposium on Automated
  Technology for Verification and Analysis}}. Springer,
  \bibinfo{pages}{181--202}.
\newblock


\bibitem[Hensel et~al\mbox{.}(2022)]%
        {hensel2022probabilistic}
\bibfield{author}{\bibinfo{person}{Christian Hensel},
  \bibinfo{person}{Sebastian Junges}, \bibinfo{person}{Joost-Pieter Katoen},
  \bibinfo{person}{Tim Quatmann}, {and} \bibinfo{person}{Matthias Volk}.}
  \bibinfo{year}{2022}\natexlab{}.
\newblock \showarticletitle{The probabilistic model checker {Storm}}.
\newblock \bibinfo{journal}{\emph{International Journal on Software Tools for
  Technology Transfer}} (\bibinfo{year}{2022}), \bibinfo{pages}{1--22}.
\newblock


\bibitem[Hensel et~al\mbox{.}(2025)]%
        {Hensel2025}
\bibfield{author}{\bibinfo{person}{Christian Hensel},
  \bibinfo{person}{Sebastian Junges}, \bibinfo{person}{Tim Quatmann}, {and}
  \bibinfo{person}{Matthias Volk}.} \bibinfo{year}{2025}\natexlab{}.
\newblock \bibinfo{booktitle}{\emph{Riding the Storm in a Probabilistic Model
  Checking Landscape}}.
\newblock \bibinfo{publisher}{Springer Nature Switzerland},
  \bibinfo{pages}{98--114}.
\newblock
\urldef\tempurl%
\url{https://doi.org/10.1007/978-3-031-75775-4\_5}
\showDOI{\tempurl}


\bibitem[Junges et~al\mbox{.}(2024)]%
        {junges2024parameter}
\bibfield{author}{\bibinfo{person}{Sebastian Junges}, \bibinfo{person}{Erika
  {\'A}brah{\'a}m}, \bibinfo{person}{Christian Hensel}, \bibinfo{person}{Nils
  Jansen}, \bibinfo{person}{Joost-Pieter Katoen}, \bibinfo{person}{Tim
  Quatmann}, {and} \bibinfo{person}{Matthias Volk}.}
  \bibinfo{year}{2024}\natexlab{}.
\newblock \showarticletitle{Parameter synthesis for {M}arkov models: covering
  the parameter space}.
\newblock \bibinfo{journal}{\emph{Formal Methods in System Design}}
  (\bibinfo{year}{2024}), \bibinfo{pages}{1--79}.
\newblock


\bibitem[Katoen(2016a)]%
        {katoen2016probabilistic}
\bibfield{author}{\bibinfo{person}{Joost-Pieter Katoen}.}
  \bibinfo{year}{2016}\natexlab{a}.
\newblock \showarticletitle{The probabilistic model checking landscape}. In
  \bibinfo{booktitle}{\emph{31st Annual ACM/IEEE Symposium on Logic in Computer
  Science}}. \bibinfo{pages}{31--45}.
\newblock


\bibitem[Katoen(2016b)]%
        {10.1145/2933575.2934574}
\bibfield{author}{\bibinfo{person}{Joost-Pieter Katoen}.}
  \bibinfo{year}{2016}\natexlab{b}.
\newblock \showarticletitle{The Probabilistic Model Checking Landscape}. In
  \bibinfo{booktitle}{\emph{31st Annual ACM/IEEE Symposium on Logic in Computer
  Science}}. \bibinfo{pages}{31–45}.
\newblock
\urldef\tempurl%
\url{https://doi.org/10.1145/2933575.2934574}
\showDOI{\tempurl}


\bibitem[Kemeny et~al\mbox{.}(1976)]%
        {kemeny-etal1976}
\bibfield{author}{\bibinfo{person}{John~G. Kemeny}, \bibinfo{person}{J.~Laurie
  Snell}, {and} \bibinfo{person}{Anthony~W. Knapp}.}
  \bibinfo{year}{1976}\natexlab{}.
\newblock \bibinfo{booktitle}{\emph{Denumerable {M}arkov Chains, 2nd edition}}.
  \bibinfo{series}{Graduate Texts in Marhematics}, Vol.~\bibinfo{volume}{40}.
\newblock \bibinfo{publisher}{Springer}.
\newblock


\bibitem[Kwiatkowska et~al\mbox{.}(2011)]%
        {KNP11}
\bibfield{author}{\bibinfo{person}{M. Kwiatkowska}, \bibinfo{person}{G.
  Norman}, {and} \bibinfo{person}{D. Parker}.} \bibinfo{year}{2011}\natexlab{}.
\newblock \showarticletitle{{PRISM} 4.0: {V}erification of Probabilistic
  Real-time Systems}. In \bibinfo{booktitle}{\emph{23rd International
  Conference on Computer Aided Verification (CAV)}}
  \emph{(\bibinfo{series}{LNCS}, Vol.~\bibinfo{volume}{6806})},
  \bibfield{editor}{\bibinfo{person}{G.~Gopalakrishnan} {and}
  \bibinfo{person}{S.~Qadeer}} (Eds.). \bibinfo{publisher}{Springer},
  \bibinfo{pages}{585--591}.
\newblock


\bibitem[Kwiatkowska et~al\mbox{.}(2017)]%
        {KNP17}
\bibfield{author}{\bibinfo{person}{M. Kwiatkowska}, \bibinfo{person}{G.
  Norman}, {and} \bibinfo{person}{D. Parker}.} \bibinfo{year}{2017}\natexlab{}.
\newblock \showarticletitle{Probabilistic Model Checking: {A}dvances and
  Applications}. In \bibinfo{booktitle}{\emph{Formal System Verification}}.
  \bibinfo{publisher}{Springer}, \bibinfo{pages}{73--121}.
\newblock


\bibitem[Kwiatkowska et~al\mbox{.}(2022)]%
        {KNP22}
\bibfield{author}{\bibinfo{person}{Marta Kwiatkowska}, \bibinfo{person}{Gethin
  Norman}, {and} \bibinfo{person}{David Parker}.}
  \bibinfo{year}{2022}\natexlab{}.
\newblock \showarticletitle{Probabilistic Model Checking and Autonomy}.
\newblock \bibinfo{journal}{\emph{Annual Review of Control, Robotics, and
  Autonomous Systems}}  \bibinfo{volume}{5} (\bibinfo{year}{2022}),
  \bibinfo{pages}{385--410}.
\newblock


\bibitem[Kwiatkowska et~al\mbox{.}(2010a)]%
        {kwiatkowska2010assume}
\bibfield{author}{\bibinfo{person}{Marta Kwiatkowska}, \bibinfo{person}{Gethin
  Norman}, \bibinfo{person}{David Parker}, {and} \bibinfo{person}{Hongyang
  Qu}.} \bibinfo{year}{2010}\natexlab{a}.
\newblock \showarticletitle{Assume-guarantee verification for probabilistic
  systems}. In \bibinfo{booktitle}{\emph{16th International Conference on Tools
  and Algorithms for the Construction and Analysis of Systems (TACAS)}}.
  Springer, \bibinfo{pages}{23--37}.
\newblock


\bibitem[Kwiatkowska et~al\mbox{.}(2010b)]%
        {KNPQ10}
\bibfield{author}{\bibinfo{person}{M. Kwiatkowska}, \bibinfo{person}{G.
  Norman}, \bibinfo{person}{D. Parker}, {and} \bibinfo{person}{H. Qu}.}
  \bibinfo{year}{2010}\natexlab{b}.
\newblock \showarticletitle{Assume-Guarantee Verification for Probabilistic
  Systems}. In \bibinfo{booktitle}{\emph{Proc. 16th International Conference on
  Tools and Algorithms for the Construction and Analysis of Systems
  (TACAS'10)}} \emph{(\bibinfo{series}{LNCS}, Vol.~\bibinfo{volume}{6105})},
  \bibfield{editor}{\bibinfo{person}{J.~Esparza} {and}
  \bibinfo{person}{R.~Majumdar}} (Eds.). \bibinfo{publisher}{Springer},
  \bibinfo{pages}{23--37}.
\newblock


\bibitem[Kwiatkowska et~al\mbox{.}(2019)]%
        {KNPS19}
\bibfield{author}{\bibinfo{person}{M. Kwiatkowska}, \bibinfo{person}{G.
  Norman}, \bibinfo{person}{D. Parker}, {and} \bibinfo{person}{G. Santos}.}
  \bibinfo{year}{2019}\natexlab{}.
\newblock \showarticletitle{Equilibria-based Probabilistic Model Checking for
  Concurrent Stochastic Games}. In \bibinfo{booktitle}{\emph{Proc. 23rd
  International Symposium on Formal Methods (FM'19)}}
  \emph{(\bibinfo{series}{LNCS}, Vol.~\bibinfo{volume}{11800})}.
  \bibinfo{publisher}{Springer}, \bibinfo{pages}{298--315}.
\newblock


\bibitem[Kwiatkowska et~al\mbox{.}(2018)]%
        {Kwiatkowska18PW}
\bibfield{author}{\bibinfo{person}{Marta Kwiatkowska}, \bibinfo{person}{David
  Parker}, {and} \bibinfo{person}{Clemens Wiltsche}.}
  \bibinfo{year}{2018}\natexlab{}.
\newblock \showarticletitle{{PRISM-games: V}erification and strategy synthesis
  for stochastic multi-player games with multiple objectives}.
\newblock \bibinfo{journal}{\emph{Int. J. Softw. Tools Technol. Transf.}}
  \bibinfo{volume}{20}, \bibinfo{number}{2} (\bibinfo{year}{2018}),
  \bibinfo{pages}{195–210}.
\newblock
\urldef\tempurl%
\url{https://doi.org/10.1007/s10009-017-0476-z}
\showDOI{\tempurl}


\bibitem[Kwiatkowska et~al\mbox{.}(2007)]%
        {KwiatkowskaNP07}
\bibfield{author}{\bibinfo{person}{Marta~Z. Kwiatkowska},
  \bibinfo{person}{Gethin Norman}, {and} \bibinfo{person}{David Parker}.}
  \bibinfo{year}{2007}\natexlab{}.
\newblock \showarticletitle{Stochastic Model Checking}. In
  \bibinfo{booktitle}{\emph{Formal Methods for Performance Evaluation, 7th
  International School on Formal Methods for the Design of Computer,
  Communication, and Software Systems, {SFM}}} \emph{(\bibinfo{series}{LNCS},
  Vol.~\bibinfo{volume}{4486})}. \bibinfo{publisher}{Springer},
  \bibinfo{pages}{220--270}.
\newblock


\bibitem[Liu and Li(2020)]%
        {liu2020compositional}
\bibfield{author}{\bibinfo{person}{Yang Liu} {and} \bibinfo{person}{Rui Li}.}
  \bibinfo{year}{2020}\natexlab{}.
\newblock \showarticletitle{Compositional stochastic model checking
  probabilistic automata via assume-guarantee reasoning}.
\newblock \bibinfo{journal}{\emph{International Journal of Networked and
  Distributed Computing}} \bibinfo{volume}{8}, \bibinfo{number}{2}
  (\bibinfo{year}{2020}), \bibinfo{pages}{94--107}.
\newblock


\bibitem[Nocedal and Wright(1999)]%
        {nocedal1999numerical}
\bibfield{author}{\bibinfo{person}{Jorge Nocedal} {and}
  \bibinfo{person}{Stephen~J Wright}.} \bibinfo{year}{1999}\natexlab{}.
\newblock \bibinfo{booktitle}{\emph{Numerical optimization}}.
\newblock \bibinfo{publisher}{Springer}.
\newblock


\bibitem[Norman et~al\mbox{.}(2017)]%
        {NPZ17}
\bibfield{author}{\bibinfo{person}{G. Norman}, \bibinfo{person}{D. Parker},
  {and} \bibinfo{person}{X. Zou}.} \bibinfo{year}{2017}\natexlab{}.
\newblock \showarticletitle{Verification and Control of Partially Observable
  Probabilistic Systems}.
\newblock \bibinfo{journal}{\emph{Real-Time Systems}} \bibinfo{volume}{53},
  \bibinfo{number}{3} (\bibinfo{year}{2017}), \bibinfo{pages}{354--402}.
\newblock


\bibitem[Organization(2021)]%
        {who2021}
\bibfield{author}{\bibinfo{person}{World~Health Organization}.}
  \bibinfo{year}{2021}\natexlab{}.
\newblock \bibinfo{title}{Ageing and Health}.
\newblock
\newblock
\urldef\tempurl%
\url{https://www.who.int/news-room/fact-sheets/detail/ageing-and-health}
\showURL{%
\tempurl}
\newblock
\shownote{Accessed: 2024-12-05}.


\bibitem[Pasareanu et~al\mbox{.}(2023)]%
        {DBLP:conf/cav/PasareanuMGYICY23}
\bibfield{author}{\bibinfo{person}{Corina~S. Pasareanu}, \bibinfo{person}{Ravi
  Mangal}, \bibinfo{person}{Divya Gopinath}, \bibinfo{person}{Sinem~Getir
  Yaman}, \bibinfo{person}{Calum Imrie}, \bibinfo{person}{Radu Calinescu},
  {and} \bibinfo{person}{Huafeng Yu}.} \bibinfo{year}{2023}\natexlab{}.
\newblock \showarticletitle{Closed-Loop Analysis of Vision-Based Autonomous
  Systems: {A} Case Study}. In \bibinfo{booktitle}{\emph{35th International
  Conference on Computer Aided Verification (CAV)}}
  \emph{(\bibinfo{series}{Lecture Notes in Computer Science},
  Vol.~\bibinfo{volume}{13964})}, \bibfield{editor}{\bibinfo{person}{Constantin
  Enea} {and} \bibinfo{person}{Akash Lal}} (Eds.).
  \bibinfo{publisher}{Springer}, \bibinfo{pages}{289--303}.
\newblock
\urldef\tempurl%
\url{https://doi.org/10.1007/978-3-031-37706-8\_15}
\showDOI{\tempurl}


\bibitem[Paterson and Calinescu(2020)]%
        {DBLP:journals/tse/PatersonC20}
\bibfield{author}{\bibinfo{person}{Colin Paterson} {and} \bibinfo{person}{Radu
  Calinescu}.} \bibinfo{year}{2020}\natexlab{}.
\newblock \showarticletitle{Observation-Enhanced {QoS} Analysis of
  Component-Based Systems}.
\newblock \bibinfo{journal}{\emph{{IEEE} Trans. Software Eng.}}
  \bibinfo{volume}{46}, \bibinfo{number}{5} (\bibinfo{year}{2020}),
  \bibinfo{pages}{526--548}.
\newblock
\urldef\tempurl%
\url{https://doi.org/10.1109/TSE.2018.2864159}
\showDOI{\tempurl}


\bibitem[Pnueli(1977)]%
        {pnueli1977temporal}
\bibfield{author}{\bibinfo{person}{Amir Pnueli}.}
  \bibinfo{year}{1977}\natexlab{}.
\newblock \showarticletitle{The temporal logic of programs}. In
  \bibinfo{booktitle}{\emph{18th Annual Symposium on Foundations of Computer
  Science}}. \bibinfo{pages}{46--57}.
\newblock


\bibitem[Powell(2007)]%
        {powell2007view}
\bibfield{author}{\bibinfo{person}{Michael~JD Powell}.}
  \bibinfo{year}{2007}\natexlab{}.
\newblock \showarticletitle{A view of algorithms for optimization without
  derivatives}.
\newblock \bibinfo{journal}{\emph{Mathematics Today-Bulletin of the Institute
  of Mathematics and its Applications}} \bibinfo{volume}{43},
  \bibinfo{number}{5} (\bibinfo{year}{2007}), \bibinfo{pages}{170--174}.
\newblock


\bibitem[Segala and Lynch(1995)]%
        {segala1995probabilistic}
\bibfield{author}{\bibinfo{person}{Roberto Segala} {and} \bibinfo{person}{Nancy
  Lynch}.} \bibinfo{year}{1995}\natexlab{}.
\newblock \showarticletitle{Probabilistic simulations for probabilistic
  processes}.
\newblock \bibinfo{journal}{\emph{Nordic Journal of Computing}}
  \bibinfo{volume}{2}, \bibinfo{number}{2} (\bibinfo{year}{1995}),
  \bibinfo{pages}{250--273}.
\newblock


\bibitem[Tarjan(1972)]%
        {Tarjan1972DepthFirstSA}
\bibfield{author}{\bibinfo{person}{Robert~Endre Tarjan}.}
  \bibinfo{year}{1972}\natexlab{}.
\newblock \showarticletitle{Depth-First Search and Linear Graph Algorithms}.
\newblock \bibinfo{journal}{\emph{SIAM J. Comput.}}  \bibinfo{volume}{1}
  (\bibinfo{year}{1972}), \bibinfo{pages}{146--160}.
\newblock
\urldef\tempurl%
\url{https://api.semanticscholar.org/CorpusID:16467262}
\showURL{%
\tempurl}


\bibitem[{ULTIMATE project}(2025)]%
        {ULTIMATE-repo-non-anonymised}
\bibfield{author}{\bibinfo{person}{{ULTIMATE project}}.}
  \bibinfo{year}{2025}\natexlab{}.
\newblock \bibinfo{title}{GitHub repository}.
\newblock
\newblock
\newblock
\shownote{\url{https://github.com/ULTIMATE-YORK/ULTIMATE}}.


\bibitem[{United Nations Population Fund (UNFPA)}(2022)]%
        {unfpa2022}
\bibfield{author}{\bibinfo{person}{{United Nations Population Fund (UNFPA)}}.}
  \bibinfo{year}{2022}\natexlab{}.
\newblock \bibinfo{title}{The State of World Population 2022: Seeing the
  Invisible}.
\newblock
\newblock
\urldef\tempurl%
\url{https://www.unfpa.org/ageing}
\showURL{%
\tempurl}
\newblock
\shownote{Accessed: 2024-12-05}.


\bibitem[Zhao et~al\mbox{.}(2020)]%
        {DBLP:conf/kbse/ZhaoCGRF20}
\bibfield{author}{\bibinfo{person}{Xingyu Zhao}, \bibinfo{person}{Radu
  Calinescu}, \bibinfo{person}{Simos Gerasimou}, {et~al\mbox{.}}}
  \bibinfo{year}{2020}\natexlab{}.
\newblock \showarticletitle{Interval Change-Point Detection for Runtime
  Probabilistic Model Checking}. In \bibinfo{booktitle}{\emph{35th {IEEE/ACM}
  International Conference on Automated Software Engineering(ASE)}}.
  \bibinfo{publisher}{{IEEE}}, \bibinfo{pages}{163--174}.
\newblock
\urldef\tempurl%
\url{https://doi.org/10.1145/3324884.3416565}
\showDOI{\tempurl}


\bibitem[Zhao et~al\mbox{.}(2024)]%
        {zhao2024bayesian}
\bibfield{author}{\bibinfo{person}{Xingyu Zhao}, \bibinfo{person}{Simos
  Gerasimou}, \bibinfo{person}{Radu Calinescu}, {et~al\mbox{.}}}
  \bibinfo{year}{2024}\natexlab{}.
\newblock \showarticletitle{Bayesian learning for the robust verification of
  autonomous robots}.
\newblock \bibinfo{journal}{\emph{Nature Comms. Eng.}} \bibinfo{volume}{3},
  \bibinfo{number}{1} (\bibinfo{year}{2024}), \bibinfo{pages}{18}.
\newblock


\end{thebibliography}
